\documentclass[aps,pra,reprint,twocolumn,superscriptaddress,amsmath,amssymb,floatfix]{revtex4-2}

\usepackage{graphicx}
\usepackage{dcolumn}
\usepackage{bm}
\usepackage{physics}
\usepackage{siunitx}
\usepackage{booktabs}
\usepackage{xcolor}
\usepackage{hyperref}

\hypersetup{
    colorlinks=true,
    linkcolor=blue,
    citecolor=blue,
    urlcolor=blue
}

\begin{document}

%\title{Optimizing Quantum Battery via Counterdiabatic driving}
\title{Geodesic Efficiency of Counterdiabatic Charging in a Quantum Battery}
\author{F. Ahmadi}
%\email{your.email@institution.edu}
\affiliation{ Department of Engineering Sciences and Physics, Buein Zahra Technical University, Buein Zahra, 
Qazvin, Iran }

\author{S. Salimi}
\affiliation{Department of Physics, University of Kurdistan, P.O.Box 66177-15175 , Sanandaj, Iran}

\begin{abstract}
We investigate an open quantum battery consisting of a resonator acting as the energy-storage unit, a frequency-tunable two-level system acting as a controllable mediator, and an external coherent  drive serving as the energy source. In the dispersive regime, the dynamics reduce to those of a driven harmonic oscillator with a tunable effective coupling. Starting from the vacuum state, the resonator remains in a coherent state under a Lindblad master equation describing photon leakage at zero temperature. We derive a counterdiabatic (CD) driving protocol that compensates for  dynamical lag and ensures that the system follows the prescribed reference trajectory. As a result, the energy stored in the resonator increases monotonically without transient oscillations, allowing the battery to reach the target ergotropy at the prescribed final time $t_f$. The CD trajectory follows a geodesic in the coherent-state phase space and achieves near-unity geodesic efficiency over a broad range of pulse durations, whereas  unassisted charging exhibits markedly lower efficiencies. Finally, we show that the CD protocol saturates the quantum speed limit for the charging process.
% We further derive a protocol-independent lower bound on the product of counterdiabatic control cost and dissipative losses, revealing an intrinsic trade-off between control effort and energy dissipation for a prescribed charging target.

\end{abstract}

\maketitle

\section{Introduction}

Energy management at the nanoscale has emerged as a fundamental challenge in quantum thermodynamics, driving the rapid development of quantum batteries (QBs) \cite{Alicki2013}. Using quantum-mechanical resources such as coherence and entanglement, QBs offer potential advantages in charging power, efficiency, and capacity over their classical counterparts \cite{Kamin12020, Kamin22020,Kamin2023,Kamin2024}. A typical quantum battery configuration consists of a  charger coupled to an energy storage system. While various physical platforms have been explored \cite{Le2018,  Zhang2019, Lu2025}, cavity-based configurations have also been proposed for quantum battery  implementations \cite{Ferraro2018, Shaghagh2022, Salvia2023, Haseli2024}. High-quality electromagnetic cavities can support multiple modes for high-capacity energy storage while enabling long-term energy retention by isolating the stored energy from rapid environmental dissipation \cite{Li2025}.

Efficient energy storage and extraction in these systems require precise quantum control. Adiabatic processes, which preserve the nondegenerate instantaneous eigenstates of a time-dependent Hamiltonian, are widely utilized due to their inherent robustness against parameter fluctuations \cite{Kato1950}. However, the fundamental requirement for slow evolution makes adiabatic protocols highly susceptible to environmentally induced decoherence.

To address the limitations of slow adiabatic evolution, shortcuts to adiabaticity (STA) have been developed to find fast dynamical trajectories that connect the initial and final states of a reference adiabatic protocol \cite{Chen2010}. Among various STA techniques,  CD driving adds a specifically designed auxiliary control field to the reference Hamiltonian to explicitly suppress unwanted diabatic transitions \cite{Demirplak2003, Berry2009}. This adiabatic-following property can enhance the robustness of CD driving against control errors and suitable for applications ranging from fast holonomic gates to efficient quantum heat engines. While STA protocols were originally developed for closed quantum systems, their application has  been extended to open system dynamics to mitigate dissipation during finite-time operations \cite{Zho2017}. 

While previous studies have demonstrated the effectiveness of CD driving in accelerating state preparation in driven–dissipative bosonic systems \cite{nature}, the requirements of quantum-battery charging extend beyond accurate and rapid state preparation. In particular, accelerating the charging process requires additional control resources, whereas prolonged charging increases cumulative energy dissipation. Consequently, achieving a prescribed target state within a shorter duration does not necessarily imply resource-efficient energy storage. Although switchable coherent-state quantum batteries have been investigated \cite{Li2025} and the fundamental trade-off between evolution speed and counterdiabatic control cost has been established \cite{campbell_2017}, the interplay between geometric efficiency, charging speed, control fluence, and dissipative energy losses warrants further investigation. Addressing this interplay is essential for identifying charging protocols that balance rapid energy storage with experimentally feasible control requirements.
In this work, we investigate the charging dynamics of an open quantum battery 
modeled as a bosonic cavity mode coupled to a driven two-level system that acts as 
the charger. Photon-loss dissipation is described using a Lindblad master equation. Working in a dispersive regime that admits an 
effective Hamiltonian, we first characterize the baseline charging performance 
under adiabatic driving and identify the principal limitations imposed by 
environmental decoherence. We then develop a customized CD protocol 
that suppresses unwanted diabatic transitions during finite-time operation and 
introduce a fluence-based cost function to optimize the CD field amplitude while 
respecting the constraints of the dispersive approximation.

We demonstrate that the CD protocol substantially enhances stored energy and 
ergotropy relative to adiabatic evolution. The state-space trajectory under CD 
driving closely follows the geodesic path, as quantified by the geodesic 
efficiency $\eta_{\rm geo}$.
The numerical  results are presented for a circuit QED platform, which provides experimentally 
accessible parameter values.  
%اضافه شد برای باند
In addition to investigating the charging performance and control fluence, we establish an analytical lower bound connecting the counterdiabatic control cost with dissipative losses for a fixed target energy, providing a general constraint on the achievable charging protocols within our model.
%اضافه شد برای باند
The remainder of the paper is organized as follows. 
In Sec.~\ref{model},  we introduce the  model and
derive the effective dispersive Hamiltonian that governs the charging
dynamics.  Sec.~\ref{sec:energy} describes  the evolution of the reduced battery density matrix  using the Lindblad master equation.
In Sec.~\ref{sec:protocols}, we analyze adiabatic charging in the open system 
and show how the CD protocol optimizes the charging of the QB and 
its stored energy.  We derive the geodesic efficiency and cost function of the CD protocol.  Although the numerical results are 
presented for a circuit QED setting, 
Sec.~\ref{sec_experimental} introduces two additional experimental platforms 
to which the protocol can be applied. We conclude in Sec.~\ref{sec_conclusion} with a discussion of practical
implications and directions for future work.

%++++++++++++++++++++++++++++++++++++++++++++++++++++++++++++++++++++++++
%اگر پالس ثابت باشد مسیر کوتاه را طی نمی کند اگر آدیاباتیک باشد طی می کند. این توضیح داده شود ولی به شکل نیازی نیست.
%در واقع طراحی این مدل کوتاهترین مسیر را برای کانتردیاباتیک به ما می دهد
%++++++++++++++++++++++++++++++++++++++++++++++++++++++++++++++++++++++++
%++++++++++++++++++++++++++++++++++++++++++++++++++++++++++++++++++++++++
\section{Physical Model and Effective Hamiltonian}
\label{model}
We consider an open quantum battery configuration that can be implemented on various experimental platforms. The setup comprises a high-quality  bosonic resonator acting as the quantum battery (the energy storage unit), a frequency-tunable two-level system (qubit)  serving as a controllable mediator that governs the energy transfer, and a coherent external driving field serving as the energy source.

%We consider a switchable open quantum battery implemented within a circuit quantum electrodynamics (cQED) architecture. 
%The setup comprises three primary elements: a high-$Q$ superconducting transmission line resonator serving as the quantum battery, an external classical microwave drive acting as the energy source, and a frequency-tunable superconducting transmon qubit acting as a controllable mediator, or charger.

\begin{figure}[t]
    \centering
    \includegraphics[width=0.9\linewidth]{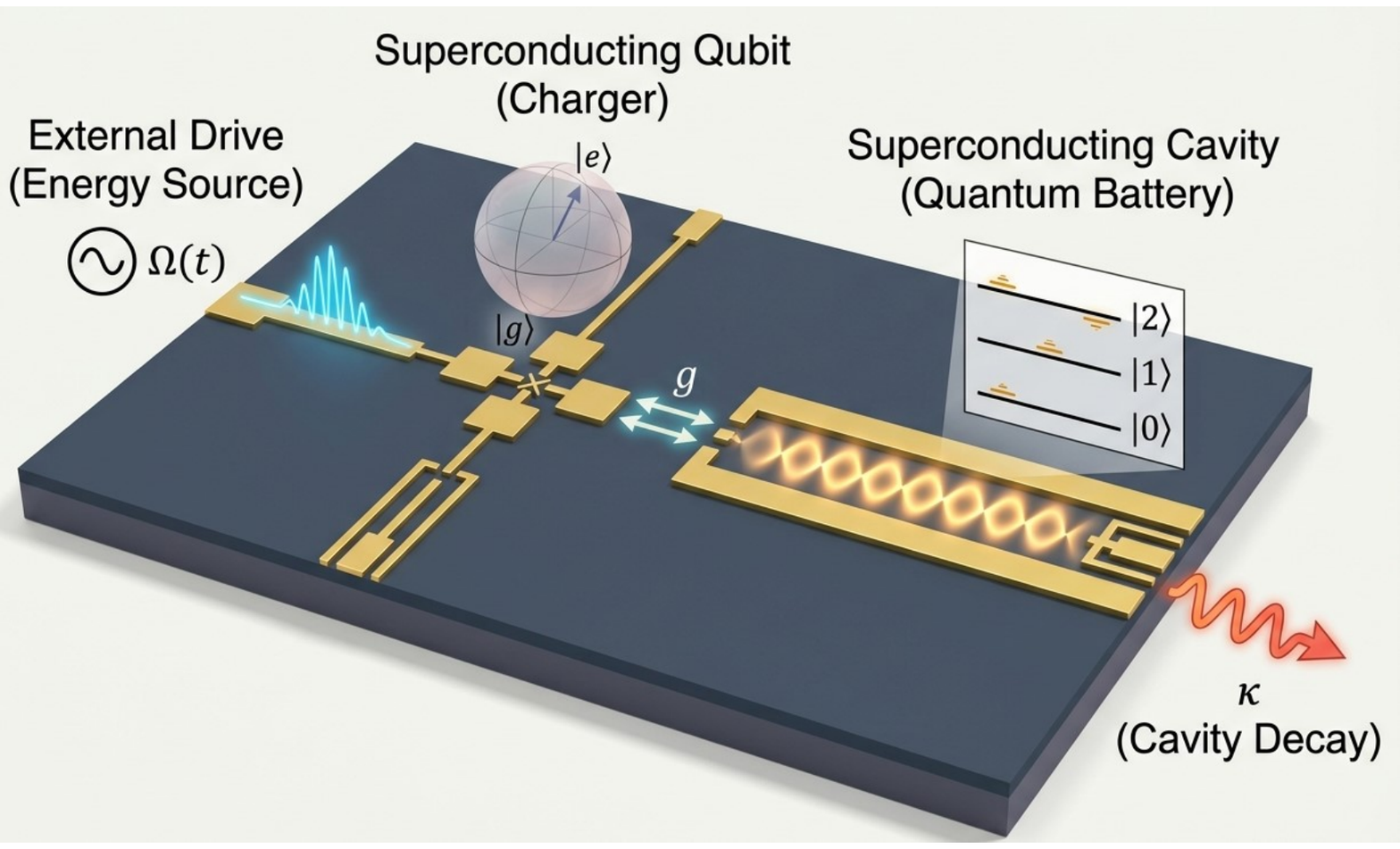}
 \caption{Schematic of a superconducting quantum battery architecture. An external drive at frequency $\omega_d$ and strength $\Omega(t)$ targets a transmon qubit charger (frequency $\omega_q$), which is coupled with strength $g$ to a superconducting transmission line cavity (frequency $\omega_a$). The cavity is subject to energy dissipation at a rate $\kappa$.}
   
    \label{fig:dissipation_analysis}
\end{figure}

Setting $\hbar = 1$,  the coupled qubit-cavity system is described by the standard Jaynes-Cummings Hamiltonian supplemented with a time-dependent coherent drive applied directly to the qubit, yielding
\begin{eqnarray}
    \hat{H}(t) &=& \frac{\omega_q}{2}\hat{\sigma}_z + \omega_a \hat{a}^\dagger \hat{a} + g \left(\hat{a}^\dagger \hat{\sigma}^- + \hat{a}\hat{\sigma}^+\right) \\ \nonumber
&+& \Omega(t)\left(e^{i\omega_d t}\hat{\sigma}^- + e^{-i\omega_d t}\hat{\sigma}^+\right).
    \label{eq:bare_H}
\end{eqnarray}
Here, $\omega_q$ denotes the tunable transition frequency of the qubit, $\omega_a$ is the fundamental resonance frequency of the storage cavity, and $g$ represents the  qubit-cavity coupling strength. The classical drive is characterized by a carrier frequency $\omega_d$ and a time-dependent amplitude $\Omega(t)$. The standard Pauli operators for the two-level qubit subspace are denoted by $\hat{\sigma}^+ = |e\rangle\langle g|$, $\hat{\sigma}^- = |g\rangle\langle e|$, and $\hat{\sigma}_z = |e\rangle\langle e| - |g\rangle\langle g|$.

To eliminate the explicit time dependence of the carrier oscillations, we transform the system into a frame rotating at the drive frequency $\omega_d$ via the unitary operator $\hat{U}(t) = \exp[i\omega_d t (\hat{a}^\dagger \hat{a} + \hat{\sigma}_z/2)]$. In this rotating frame, the Hamiltonian becomes
\begin{equation}
    \hat{H}'(t) = \frac{\Delta_q}{2}\hat{\sigma}_z + \Delta_a \hat{a}^\dagger \hat{a} + g \left(\hat{a}^\dagger \hat{\sigma}^- + \hat{a}\hat{\sigma}^+\right) + \Omega(t)(\hat{\sigma}^-+\hat{\sigma}^+).
    \label{eq:rotating_H}
\end{equation}
The qubit and cavity detunings relative to the drive frequency are defined as $\Delta_q \equiv \omega_q - \omega_d$ and $\Delta_a \equiv \omega_a - \omega_d$, respectively, while the mutual qubit-cavity detuning is $\Delta \equiv \omega_q - \omega_a = \Delta_q - \Delta_a$.

We operate the system in the dispersive regime, where direct real excitations of the qubit are strongly suppressed. This requires the conditions
\begin{equation}
    \frac{g}{|\Delta|} \ll 1, \quad \frac{|\Omega(t)|}{|\Delta_q|} \ll 1.
    \label{eq:dispersive_conditions}
\end{equation}
Under these constraints, the mediator qubit remains predominantly in its ground state $|g\rangle$, facilitating energy transfer from the classical drive to the cavity via virtual excitations. Applying a standard Fr\"ohlich-Nakajima (Schrieffer-Wolff) transformation up to second order in $g/|\Delta|$ and $\Omega/|\Delta_q|$ eliminates the first-order interaction terms \cite{Li2025}. We then project the dynamics onto the qubit ground-state subspace ($\langle \hat{\sigma}_z \rangle = -1$, $\langle \hat{\sigma}^+ \hat{\sigma}^- \rangle = 0$), invoke the rotating-wave approximation to neglect fast-oscillating terms, and discard irrelevant constant energy shifts. The effective Hamiltonian governing the cavity battery thereby simplifies to that of a driven displaced harmonic oscillator, expressed as
\begin{equation}
    \hat{H}_{\rm eff}(t) = \omega_{\rm eff} \hat{a}^\dagger \hat{a} - \lambda \Omega(t)\left(\hat{a}^\dagger + \hat{a}\right),
    \label{eq:Heff_final}
\end{equation}
where the effective cavity detuning $\omega_{\rm eff}$ and the tunable driving coupling $\lambda$ are defined as 
\begin{align}
    \omega_{\rm eff} &= \Delta_a - \frac{g^2}{\Delta}, \label{eq:omega_eff} \\
    \lambda &= g\left(\frac{1}{\Delta_q} + \frac{1}{\Delta}\right) \label{eq:lambda}
\end{align}
(see Appendix \ref{app:derivation_effective} for details).  Equation~\eqref{eq:Heff_final} demonstrates that the dispersive qubit-cavity interaction maps the driven composite  system onto an effective cavity model subject to a classical drive with strength $\lambda\Omega(t)$ and a Stark-shifted frequency $\omega_{\rm eff}$. Because $\lambda$ depends on the qubit detuning $\Delta_q$, the effective coupling can be dynamically tuned or completely switched off. This provides a controllable gateway to charge the battery on demand while isolating it to prevent energy leakage during the storage phase.

%Finally, the validity of the dispersive Hamiltonian in Eq.~\eqref{eq:Heff_final} requires the intracavity photon population to remain well below the critical photon number \cite{Blais2004,Blais2021}, written as
%\begin{equation}
%    \langle \hat{a}^\dagger \hat{a} \rangle \ll n_{\rm crit} \equiv \frac{\Delta^2}{4g^2}.
%    \label{eq:ncrit}
%\end{equation}
%Beyond this threshold, transmon nonlinearity, higher-order multiphoton transitions, and the breakdown of the dispersive approximation can no longer be neglected.

%++++++++++++++++++++++++++++++++++++++++++++++++++++++
%++++++++++++++++++++++++++++++++++++++++++++++++++++++
%+++++++++++++++++++++++++++++++++++++++++++++++++++++++++++++++++++++++++++++++++++++++
\section{Charging Dynamics }\label{sec:energy}

\subsection{Open-System Dynamics and State Purity}
\label{subsec:open_system_dynamics}

To model energy dissipation during the charging process, we account for photon leakage through the cavity  into the  environment. Assuming a zero-temperature Markovian bath ($\bar{n}_{\rm th} \approx 0$), 
%at typical dilution refrigerator temperatures ($T \sim 10\text{ mK}$), thermal fluctuations are negligible ($\bar{n}_{\rm th} \approx 0$). Thus, 
the evolution of the reduced battery density matrix $\hat{\rho}(t)$ is governed by the Lindblad master equation
\begin{equation}
    \frac{d\hat{\rho}(t)}{dt} = -i \left[\hat{H}_{\rm eff}(t), \hat{\rho}(t)\right] + \kappa \mathcal{D}[\hat{a}]\hat{\rho}(t),
    \label{eq:master_eq}
\end{equation}
where $\kappa$ is the cavity decay rate and $\mathcal{D}[\hat{a}]\hat{\rho} = \hat{a}\hat{\rho}\hat{a}^\dagger - \frac{1}{2}\{\hat{a}^\dagger\hat{a}, \hat{\rho}\}$ is the standard dissipator.

Because the effective Hamiltonian $\hat{H}_{\rm eff}(t)$ remains linear in the bosonic field operators, an initially empty cavity, $\hat{\rho}(0) = |0\rangle\langle 0|$, evolves strictly as a pure coherent state $\hat{\rho}(t) = |\alpha(t)\rangle\langle\alpha(t)|$ throughout the open-system dynamics (see Appendix~\ref{app:coherent_state_proof} for a detailed proof). The complex amplitude $\alpha(t) = \langle \hat{a}(t) \rangle$ satisfies the equation of motion
\begin{equation}
    \frac{d\alpha(t)}{dt} = -\gamma_{\rm eff}\alpha(t) + i\lambda\Omega(t),
    \label{eq:eom_alpha}
\end{equation}
where 
\begin{equation}
  \gamma_{\rm eff} = i\omega_{\rm eff} + \kappa/2.
    \label{eq:gamma_eff}
\end{equation}
The exact analytical solution of Eq.~\eqref{eq:eom_alpha} is given by
\begin{equation}
    \alpha(t) = i\lambda \int_{0}^{t} e^{-\gamma_{\rm eff}(t-t')} \Omega(t') \, dt'.
    \label{eq:alpha_exact_solution}
\end{equation}
Thus, specifying the driving protocol $\Omega(t)$ determines the coherent amplitude $\alpha(t)$ through Eq.~\eqref{eq:alpha_exact_solution}, and the choice of $\Omega(t)$ directly influences the resulting cavity amplitude and, consequently, the energy stored in the quantum battery as discussed in the following sections.
%---------------------------------------------------------------
\subsection{ Energy, Power, and Ergotropy}

The battery's stored  energy  is evaluated with respect to the bare cavity Hamiltonian $\hat{H}_B = \omega_a \hat{a}^\dagger \hat{a}$, 
\begin{equation}
    E(t) = \text{Tr}\left[ \hat{H}_B \hat{\rho}(t) \right] = \omega_a |\alpha(t)|^2 .
    \label{eq:stored_energy}
\end{equation}
The instantaneous charging power $P(t) \equiv dE(t)/dt$ can be separated into a drive-induced and a dissipative contributions, 
\begin{equation}
    P(t) = P_{\mathrm{in}}(t) - P_{\mathrm{diss}}(t) = 2\omega_a \lambda \Omega(t) \mathrm{Im}[\alpha(t)] - \kappa E(t) .
    \label{eq:power_balance}
\end{equation}
Here,  $P_{\mathrm{in}}(t)$ is the power delivered to the cavity mode by the driving field,that is, the contribution associated with the driving field $\Omega(t)$, and  $P_{\mathrm{diss}}(t)$ is the photon-leakage rate (see the details in Appendix~\ref{app:power_derivation}).  \\
%For a total charging duration $t_f$, the total input energy supplied by the drive satisfies $E_{\mathrm{in}} = \int_0^{t_f} P_{\mathrm{in}}(t)\,dt = E(t_f) + E_{\mathrm{diss}}$, where $E_{\mathrm{diss}} = \kappa \int_0^{t_f} E(t)\,dt$ represents the total energy leaked into the environment (see the details in Appendix~\ref{app:power_derivation}). Consequently, in the presence of dissipation, the fraction of injected work retained in the cavity mode is strictly less than unity, $E(t_f)/[E(t_f) + E_{\mathrm{diss}}] < 1$, fully satisfying the first law of thermodynamics.

The ergotropy $\mathcal{E}(t)$, which quantifies the maximum work extractable from the battery through cyclic unitary operations, is defined as~\cite{Allahverdyan2004}
\begin{equation}
    \mathcal{E}(t) = E(t) - \min_{\hat{U}} \text{Tr}\left[ \hat{H}_B \hat{U}\hat{\rho}(t)\hat{U}^\dagger \right] = E(t) - E_{\hat{\sigma}_\rho}
    \label{eq:ergotropy_def}
\end{equation}
where $\hat{\sigma}_\rho$ is the passive state associated with $\hat{\rho}(t)$. Because the system density matrix $\hat{\rho}(t) = |\alpha(t)\rangle\langle\alpha(t)|$ remains perfectly pure, its unique passive state is  the vacuum state $\hat{\sigma}_\rho = |0\rangle\langle 0|$, yielding $E_{\hat{\sigma}_\rho} = 0$. Therefore, we find
\begin{equation}
    \mathcal{E}(t) = E(t) = \omega_a |\alpha(t)|^2,
    \label{eq:ergotropy_pure}
\end{equation}
which confirms that the total energy stored in the cavity at time $t_f$ is coherent and fully extractable as useful quantum work. Thus, maximizing the extractable work $\mathcal{E}(t_f)$ is formally equivalent to maximizing the coherent photon number $|\alpha(t_f)|^2$.
%++++++++++++++++++++++++++++++++++++++++++++++++++++++++++++++++
%++++++++++++++++++++++++++++++++++++++++++++++++++++++++++++++++
%++++++++++++++++++++++++++++++++++++++++++++++++++++++++++++++++
\section{Charging Protocols }
\label{sec:protocols}

\subsection{ Constant Driving }\label{subsec:constant}

\begin{figure}[htb!!]
    \centering
    \includegraphics[width=1\linewidth]{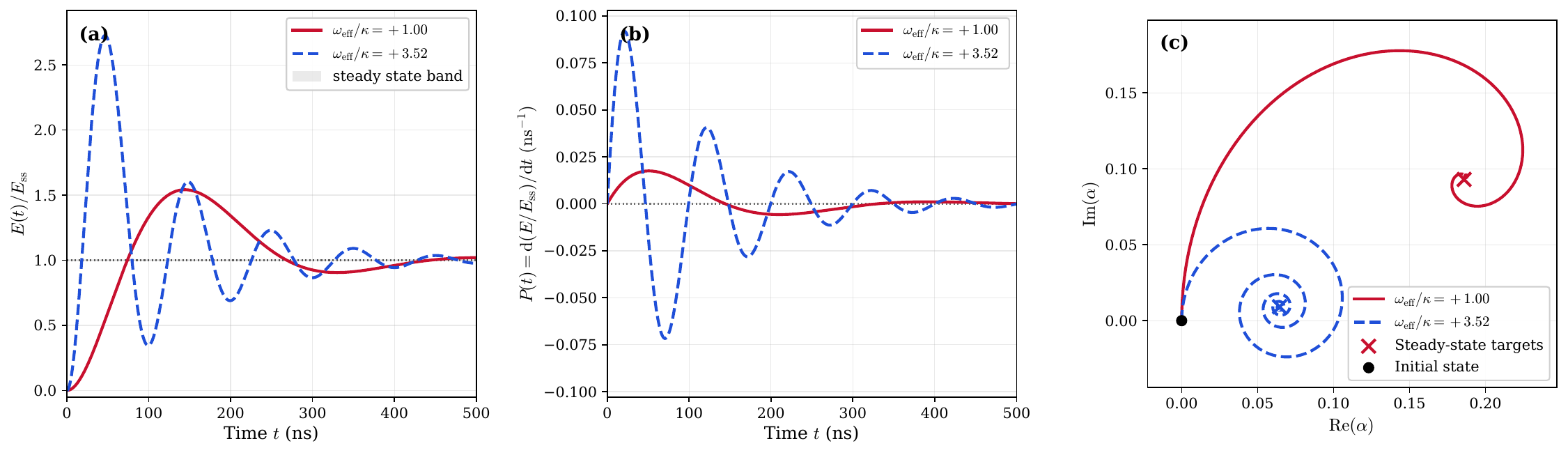}
   \caption{\textbf{
Open-cavity battery dynamics under a constant coherent drive $\Omega_0$  compared for two different   ratios $\omega_{\text{eff}}/\kappa$. }
(a) Normalized stored energy $E(t)/E_{\text{ss}}$ as a function of time $t$, where the shaded horizontal band indicates the $1\%$ neighborhood around the steady-state target energy. 
(b) Instantaneous charging power $P(t) = \mathrm{d}(E/E_{\text{ss}})/\mathrm{d}t$.
(c) Phase-space trajectories of $\alpha(t)$ illustrating the spiral convergence toward their respective steady-state targets $\alpha_{\text{ss}}$ (marked by crosses), with an initial state at the origin (filled circle). 
The fixed physical parameters for the red curve are given in Table~\ref{tab:parameters}. For the dashed blue curve, only the drive frequency is  $\omega_d/2\pi = 5.2~\mathrm{GHz}$.
}
    \label{fig:dynamic_const_charge}
\end{figure}

As a baseline for  the cavity battery charging dynamics, we first examine a square drive pulse of constant amplitude,  $\Omega_0$,  applied over the charging interval,
\begin{equation}
    \Omega(t) = 
    \begin{cases}
      \Omega_0 & 0 \le t \le t_f, \\
        0 & \text{otherwise,}
    \end{cases}
    \label{eq:constant_drive_def}
\end{equation}
where the pulse is turned on instantaneously at $t=0$ and switched off at $t=t_f$. Solving Eq.~\eqref{eq:eom_alpha} for  the cavity amplitude $\alpha(t)$, subject to  the initial vacuum condition $\alpha(0) = 0$, gives
\begin{equation}
    \alpha(t) = \frac{i\lambda\Omega_0}{\gamma_{\rm eff}} \left( 1 - e^{-\gamma_{\rm eff}t} \right).
    \label{eq:alpha_constant_exact}
\end{equation}
Using Eq.~\eqref{eq:stored_energy}, the instantaneous  energy stored  in the quantum battery is given by
\begin{equation} \label{eq:energy_constant_exact}
    E(t) = \frac{\omega_a \lambda^2 \Omega_0^2}{|\gamma_{\rm eff}|^2 } \left[ 1 + e^{-\kappa t} - 2 e^{-\kappa t/2} \cos(\omega_{\rm eff} t) \right].
\end{equation}
As illustrated in Fig.~\ref{fig:dynamic_const_charge}(a), the sudden onset of the driving field induces an underdamped oscillatory response in the normalized stored energy $E(t)/E_{\text{ss}}$ around the steady-state value, $E_{\text{ss}}=\frac{\omega_a \lambda^2 \Omega_0^2}{|\gamma_{\rm eff}|^2 }$. 
%transient dynamics در اینجا به معنی «رفتار سامانه در بازهٔ زمانی اولیه پس از روشن‌شدن ناگهانی درایو و پیش از رسیدن به حالت پایا» است.
These transient dynamics are governed by coherent oscillations at the effective frequency $\omega_{\text{eff}}$, with a decay timescale determined by the cavity decay rate $\kappa$. 
Comparing the two drive frequencies $\omega_d/(2\pi) = 5.25712~\text{GHz}$ ($\omega_{\text{eff}}/\kappa \approx 1.0$) and $\omega_d/(2\pi) = 5.25~\text{GHz}$ ($\omega_{\text{eff}}/\kappa \approx 3.5$) demonstrates how the dynamical regime shifts from moderately damped to significantly more oscillatory, exhibiting pronounced overshoots. In the weakly dissipative regime ($\kappa \ll \omega_{\rm eff}$), the final stored energy behaves approximately as $E(t_f) \approx  \frac{4\omega_a \lambda^2 \Omega_0^2 }{ \omega_{\rm eff}^2} \sin^2(\frac{\omega_{\rm eff} t_f }{ 2})$. Consequently, the final stored energy becomes highly sensitive to small timing errors, which can cause it to fluctuate between nearly zero and its maximum value.
%%«این نتیجه نشان می‌دهد که حتی نوسان‌ها یا خطاهای بسیار کوچک در زمان‌بندی اعمال میدان می‌توانند باعث تغییرات شدیدی در انرژی ذخیره‌شده شوند؛ به‌گونه‌ای که انرژی ممکن است بین مقدار تقریباً صفر و بیشینهٔ خود نوسان کند.»

The effects of these nonadiabatic transients on energy transfer are further evidenced by the normalized instantaneous charging power $P(t) = \mathrm{d}(E/E_{\text{ss}})/\mathrm{d}t$, displayed in Fig.~\ref{fig:dynamic_const_charge}(b). 
During the initial response, the power oscillates and periodically dips into negative values ($P(t) < 0$). 
Negative instantaneous power indicates temporary coherent energy backflow from the cavity mode into the driving field, accompanied by inherent cavity dissipation, prior to eventual relaxation toward the steady state. 

The underlying phase-space trajectories of $\alpha(t)$, shown in Fig.~\ref{fig:dynamic_const_charge}(c), reveal extended spiral trajectories around their respective steady-state target amplitudes $\alpha_{\text{ss}}$ (marked by crosses) rather than a direct monotonic approach from the vacuum state. 

Furthermore, abrupt pulse switching at \(t=0\) and \(t=t_f\) introduces broad spectral components that may compromise the validity of the dispersive approximation, potentially inducing unwanted excitations of the mediator or triggering unintended resonant dynamics. For nonzero effective detuning, switching off the drive at an arbitrary intermediate time may leave the resonator in a state that deviates significantly from the desired target. Although the stored energy increases monotonically at exact effective resonance (\(\omega_{\mathrm{eff}}=0\)), the approach to the steady-state energy remains limited by the finite response time of the cavity. These limitations motivate the development of tailored driving protocols for fast and deterministic charging.

%در این قسمت اشاره به این شود که اگر کاپا نزدیک به صفر باشد عملا ما درایو نوسانی خواهیم داشت. سوال: همین درایو را در یک t_f کوتاه در نظر بگیریم پاسخ مقدار انرژی خیلی کم است. اگر کاپا خیلی کم باشد نسبت به w_eff نوسانات خیلی زیادی خواهیم داشت. مقدار منفی و مثبت w_eff هم جالب است که روی جهت حرکت اثر می گذارد. 
We consider an open quantum battery implemented within a circuit quantum electrodynamics (cQED) configuration. The system consists of three main components: a high-quality superconducting transmission-line resonator serving as the quantum battery, an external classical microwave drive acting as the energy source, and a frequency-tunable superconducting transmon qubit acting as the controllable mediator (charger). To establish a realistic physical regime for the accelerated charging protocol, Table~\ref{tab:parameters} summarizes representative experimental parameters for this cQED architecture. These values, representative of superconducting circuit-QED setups \cite{Eickbusch2022, nature}, are used in our analytical calculations and numerical simulations to ensure operation well within the dispersive regime, as defined by Eq.~\eqref{eq:dispersive_conditions}.
\begin{table}[t]
\centering
\caption{Circuit QED system parameters. The effective cavity frequency $\omega_{\rm eff}$ and coupling $\lambda$ are defined in Eq.~(\ref{eq:lambda}) % ; the dispersive regime requires $\lambda \ll 1$ and $\kappa \ll \omega_{\rm eff}$, both satisfied here ($\kappa/\omega_{\rm eff} \sim 10^{-3}$).
}
\begin{ruledtabular}
\begin{tabular}{lll}
Parameter & Symbol & Value \\
\hline
Cavity frequency & $\omega_a/2\pi$ & 5.26~GHz \\
Qubit frequency & $\omega_q/2\pi$ & 6.65~GHz \\
Drive frequency & $\omega_d/2\pi$ & 5.25712~GHz\\
Qubit–cavity coupling & $g/2\pi$ & 9.12~MHz \\
%Effective cavity frequency & $\omega_{\rm eff}$ & $1.5~{\rm rad/ns} \approx 2\pi \times 240$~MHz \\
Cavity decay rate & $\kappa/2\pi$ & $2.82$~MHz \\ % $1.0 \times 10^{-3}~{\rm rad/ns} \approx
%Qubit dephasing (3D cQED) & $\gamma/2\pi$ & $\sim 5$~kHz \\
Drive amplitude & $\Omega_0/2\pi$ & 50~MHz \\
%Effective coupling & $\lambda \Omega_0$ & 1.0~rad/ns \\
%Optimal charging time & $t_f^{\rm opt}$ & $\sim 1.2$~ns \\
%Energy decay time & $\tau_E = 1/\kappa$ & 1000~ns \\
%Amplitude decay time & $\tau_\alpha = 2/\kappa$ & 2000~ns \\
%Adiabaticity ratio & $\kappa/2\omega_{\rm eff}$ & $\sim 3.3 \times 10^{-4}$ \\
\end{tabular}
\end{ruledtabular}
\label{tab:parameters}
\end{table}
%=========================================================================
%=========================================================================
%=========================================================================

\subsection{ Adiabatic Charging  }
\label{subsec:adiabatic}

To analyze the adiabatic charging, we integrate Eq.~\eqref{eq:alpha_exact_solution} by parts,
\begin{align}
    \alpha(t) &= i\lambda \int_{0}^{t} e^{-\gamma_{\rm eff}(t-t')} \Omega(t') \, dt' \nonumber \\
    &= \frac{i\lambda}{\gamma_{\rm eff}}\left[ \Omega(t) - \Omega(0)e^{-\gamma_{\rm eff} t} - \int_{0}^{t} e^{-\gamma_{\rm eff}(t-t')} \dot{\Omega}(t') \, dt' \right].
    \label{eq:alpha_parts}
\end{align}
To quantify the adiabatic condition,  we consider the instantaneous steady-state solution of Eq.~\eqref{eq:eom_alpha} defined by $\dot{\alpha}_{\mathrm{ad}}= 0$, which gives $\alpha_{\mathrm{ad}}(t) = i\frac{ \lambda \Omega(t)} { \gamma_{\mathrm{eff}}}$. We decompose the amplitude as $\alpha(t) = \alpha_{\mathrm{ad}}(t) + \delta\alpha(t)$,  where $\delta\alpha(t)$ is the nonadiabatic deviation. Substituting this decomposition into Eq.~\eqref{eq:eom_alpha} gives $\dot{\alpha}_{\mathrm{ad}} + \dot{\delta\alpha} = -\gamma_{\mathrm{eff}} \delta\alpha$. In the near-adiabatic regime where $\dot{\delta\alpha}$ is negligible compared with $\dot{\alpha}_{\mathrm{ad}}$, the first-order deviation is $\delta\alpha(t) \approx -\frac{\dot{\alpha}_{\mathrm{ad}}}{\gamma_{\mathrm{eff}}} = -i\frac{\lambda \dot{\Omega}(t)}{\gamma_{\mathrm{eff}}^2}$. Requiring this correction to remain much smaller than the instantaneous steady-state amplitude, $|\delta\alpha(t)| \ll |\alpha_{\mathrm{ad}}(t)|$,  imposes the operational adiabatic condition
\begin{equation}
    \left| \frac{\dot{\Omega}(t)}{\Omega(t)} \right| \ll |\gamma_{\mathrm{eff}}|.
    \label{eq:adiabatic_condition}
\end{equation}
Under this condition, the integral term containing $\dot{\Omega}$ in Eq.~\eqref{eq:alpha_parts} becomes negligible to leading order. 
Given $\Omega(0)=0$, the cavity state closely follows the  instantaneous  steady state,
\begin{equation}
    \alpha_{\rm ad}(t) \approx \frac{i\lambda}{\gamma_{\rm eff}} \Omega(t) %= \frac{\lambda \Omega(t)}{\omega_{\rm eff} - i\kappa/2}.
    \label{eq:alpha_ad_instantaneous}
\end{equation}
Consequently, at the end of the charging cycle $t=t_f$, the energy stored under adiabatic driving, as defined in Eq.~\eqref{eq:stored_energy} becomes
\begin{equation}
    E_{\rm ad}(t_f) = \frac{\omega_a \lambda^2 }{|\gamma_{\rm eff}|^2 }\Omega^2(t_f).
    \label{eq:energy_ad_tf}
\end{equation}
% آیا نیازی هست که این رابطه با مقدار آن در پالس ثابت مقایسه شود؟
A crucial physical contrast emerges  when considering $\Omega(t_f)=\Omega_0$ and comparing Eq.~\eqref{eq:energy_ad_tf} with the constant-drive result. First, the smooth envelope successfully suppresses non-adiabatic transient oscillations, allowing the system to track the instantaneous steady-state trajectory in phase space without developing rapid oscillatory orbits. This yields robust stability against pulse-duration variations. Second, adiabatic charging in an open system suffers from an unavoidable competition between adiabatic fidelity and dissipation. Increasing $t_f$ improves tracking fidelity but simultaneously exposes the cavity battery to continuous environmental dissipation through  decay $\kappa$.
For longer charging duration, the  dissipated energy $E_{\rm diss} = \kappa \int_0^{t_f} E(t)\,dt$ grows considerably, severely degrading both the charging efficiency  and the average charging power $\bar{P} = E(t_f)/t_f$.
Thus, conventional adiabatic charging involves a trade-off between accurate adiabatic tracking, which favors longer charging durations, and reduced energy dissipation, which favors shorter durations.
This trade-off motivates the use of CD driving, which enables faster charging while reducing the energy lost to dissipation by shortening the charging time.

%=========================================================================
%=========================================================================
%=========================================================================
\subsection{ Counterdiabatic Charging Dynamics}\label{subsec:cd_charging_dynamics}

To overcome the trade-off inherent in open-system adiabatic charging, we employ transitionless quantum driving via counterdiabatic (CD) control~\cite{Odelin2019}. In this approach, an auxiliary control field is designed that cancels nonadiabatic excitation, thereby enforcing
adiabatic tracking for arbitrary charging times $t_f$. The instantaneous equilibrium
state associated with Eq.~(\ref{eq:eom_alpha}) defines the target amplitude
\begin{equation}
    \alpha_{\rm target}(t)
    = \frac{i\lambda}{\gamma_{\rm eff}}\,\Omega(t)
   % = \frac{\lambda}{\omega_{\rm eff} - i\kappa/2}\,\Omega(t).
    \label{eq:cd_target_state}
\end{equation}
Substituting the total drive $\Omega_{\rm tot}(t)=\Omega(t)+\Omega_{\rm cd}(t)$
into Eq.~(\ref{eq:eom_alpha}) and using Eq.~\eqref{eq:alpha_parts}  we obtain
\begin{align}
    \alpha(t)
    &= i\lambda \int_{0}^{t} e^{-\gamma_{\rm eff}(t-t')}
       \bigl[\Omega(t') + \Omega_{\rm cd}(t')\bigr]\,dt' \nonumber \\
    &= \frac{i\lambda}{\gamma_{\rm eff}}
       \Biggl[
         \Omega(t) - \Omega(0)\,e^{-\gamma_{\rm eff} t}
         - \int_{0}^{t} e^{-\gamma_{\rm eff}(t-t')}\dot{\Omega}(t')\,dt'   \Biggr] \nonumber \\
    &     + i\lambda \int_{0}^{t} e^{-\gamma_{\rm eff}(t-t')}\Omega_{\rm cd}(t')\,dt'
     .
    \label{eq:alpha_parts2}
\end{align}
Using $\Omega(0)=0$ and imposing exact tracking condition $\alpha(t)=\alpha_{\rm target}(t)$ for all $t$, 
the two remaining integrals balance, giving
\begin{equation}
    \Omega_{\rm cd}(t) = \frac{\dot{\Omega}(t)}{\gamma_{\rm eff}}.
    \label{eq:cd_field_exact}
\end{equation}
%%%%%اضافه شد
It is important to distinguish the present CD protocol from conventional transitionless driving in closed quantum systems. Whereas conventional counterdiabatic driving suppresses nonadiabatic transitions between instantaneous eigenstates of a time-dependent Hamiltonian, the present open-system approach aims to maintain the system along the instantaneous steady-state trajectory of the reference Lindblad dynamics. For the driven–dissipative cavity considered here, this steady state is a coherent state characterized by the amplitude \(\alpha_{\mathrm{target}}(t)=i\lambda\Omega(t)/\gamma_{\mathrm{eff}}\). Under a time-dependent reference drive, the cavity generally exhibits a dynamical lag relative to this moving steady state. The auxiliary field derived in Eq.~\eqref{eq:cd_field_exact} compensates for this lag and enforces exact tracking of the prescribed coherent-state trajectory for any charging duration within the regime of validity of the effective model. Therefore, the counterdiabatic protocol eliminates deviations from the instantaneous reference steady state rather than suppressing the intrinsic cavity dissipation, which remains governed by the decay rate \(\kappa\).
%اضافه شد
Thus, for an arbitrary reference drive $\Omega(t)$, the instantaneous equilibrium state $ \alpha_{\rm target}(t)$ is always the exact dynamic solution of Eq.~(\ref{eq:eom_alpha}).

Expanding $\Omega_{\mathrm{CD}}(t)$ into its in-phase and quadrature components gives
\begin{eqnarray}     \label{eq:re_im_cd}
    \mathrm{Re}[\Omega_{\mathrm{CD}}(t)] &=& \dot{\Omega}(t) \, \frac{\kappa/2}{\left(\frac{\kappa}{2}\right)^2 + \omega_{\mathrm{eff}}^2}, \quad \\ \nonumber
    \mathrm{Im}[\Omega_{\mathrm{CD}}(t)] &= &-\dot{\Omega}(t) \, \frac{\omega_{\mathrm{eff}}}{\left(\frac{\kappa}{2}\right)^2 + \omega_{\mathrm{eff}}^2}.
\end{eqnarray}
%%%%%اضافه شد
The complex CD field requires both in-phase and quadrature control to achieve exact tracking of the instantaneous steady state. As shown in Eq. ~\eqref{eq:re_im_cd}, the imaginary component is associated with the coherent dynamics governed by the effective cavity frequency \(\omega_{\mathrm{eff}}\), providing the quadrature correction needed to compensate for the dynamical lag induced by the time-dependent reference drive. In contrast, the real component is proportional to the cavity decay rate \(\kappa\) and accounts for the damping-dependent contribution to the required control. In the closed-system limit (\(\kappa\rightarrow0\)), the real component vanishes, while the imaginary component reduces to the conventional counterdiabatic correction, \(\operatorname{Im}[\Omega_{\mathrm{CD}}]=-\dot{\Omega}/\omega_{\mathrm{eff}}\). Therefore, the two components jointly ensure exact tracking of the prescribed coherent-state trajectory in the open system, without eliminating the intrinsic cavity dissipation. In a circuit-QED implementation, these components can be realized through in-phase and quadrature modulation of the microwave field applied to the qubit. The auxiliary CD field can also be derived directly from the Lindblad dynamics (see Appendix.~\ref{app:liouvillian_cd}).

\begin{figure}[t!]
    \centering
    \includegraphics[width=\linewidth]{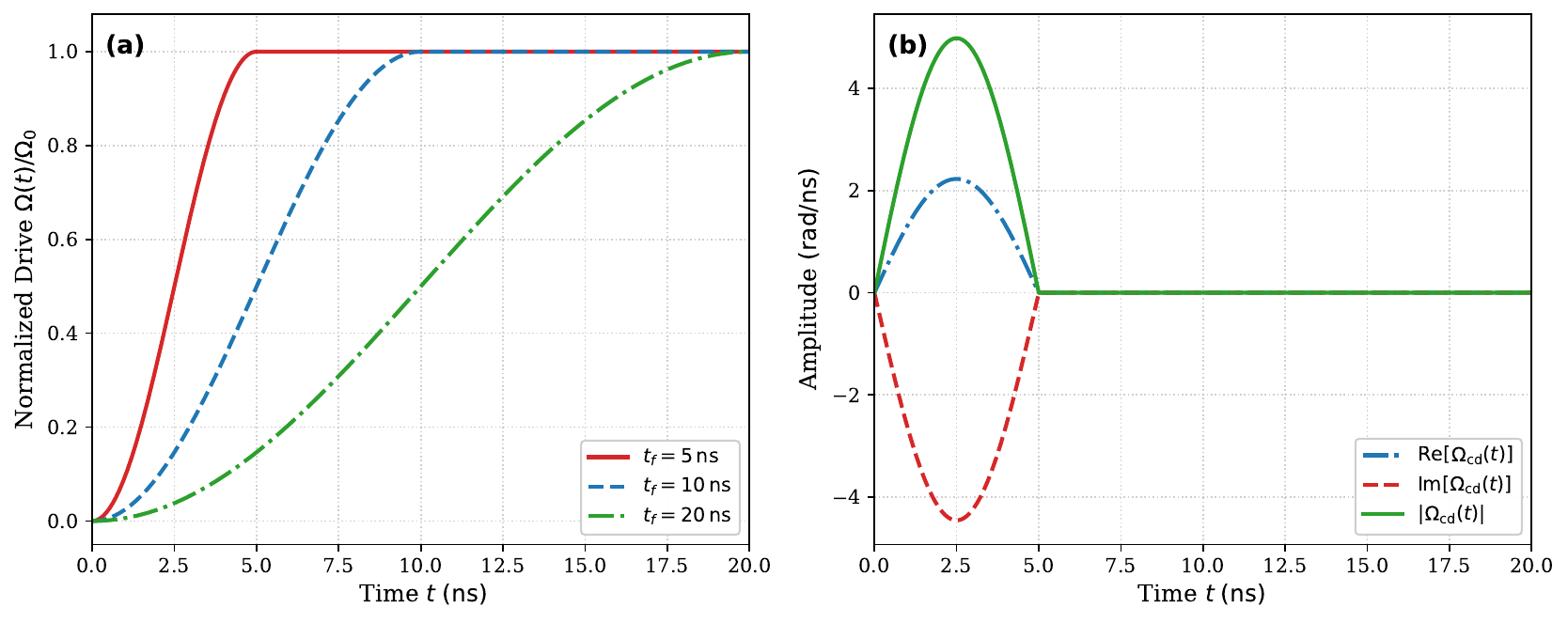}
  \caption{\textbf{Control drive pulse profiles and CD components.} (a) Normalized smooth pulse envelopes $\Omega(t)/\Omega_0$ for various charging durations $t_f = 5$, $10$, and $20\text{ ns}$. (b) Real part, imaginary part, and absolute magnitude of the CD auxiliary drive $\Omega_{\mathrm{CD}}(t)$ evaluated for $t_f = 5\text{ ns}$. The fixed physical parameters  are given in Table~\ref{tab:parameters}.}
       \label{fig:pulse_profiles}
\end{figure}

%%%%%اضافه شد
%In modern 3D circuit QED architectures ($\kappa/2 \approx 2\pi \times 5~\text{kHz}$ versus $\omega_{\mathrm{eff}} \approx 2\pi \times 1.5~\text{GHz}$), the ratio between components is suppressed by a factor of $\kappa / 2\omega_{\mathrm{eff}} \approx 3.3 \times 10^{-6}$. 
%Consequently,  $\mathrm{Re}[\Omega_{\mathrm{CD}}(t)]$ 
% is negligible, confirming that the CD correction operates via quadrature modulation, which is readily achievable using standard pulse-shaping techniques in modern quantum control experiments. In the other words,  in the weak-dissipiation limit ($\kappa \ll \omega_{\rm eff}$), Eq. (\ref {eq:cd_field_exact}) reduces to $\Omega_{\rm cd}(t) \approx -i \dot{\Omega}(t)/\omega_{\rm eff} = [\dot{\Omega}(t)/\omega_{\rm eff}]\, e^{-i\pi/2}$, indicating a precise $\pi/2$ phase shift relative to the primary envelope $\Omega(t)$.
As shown in Fig.~\ref{fig:pulse_profiles}, we  apply  a $\sin^2$-shaped pulse for the reference driving $\Omega(t)$~\cite{nature}
\begin{equation}
    \Omega(t) = \Omega_0 \sin^2\left(\frac{\pi t}{2 t_f}\right), \quad 0 \le t \le t_f,
    \label{eq:adiabatic_pulse_def}
\end{equation}
which satisfies smooth boundary conditions at both the onset and the target charging time, namely $\Omega(0)=0$, $\dot{\Omega}(0)=0$, and $\dot{\Omega}(t_f)=0$ with $\Omega(t_f)=\Omega_0$. 

For the $\sin^2$-shaped reference pulse, the auxiliary CD field defined in Eq.~\eqref{eq:cd_field_exact} takes the form
\begin{equation}
    \Omega_{\rm cd}(t) = \, \frac{\pi \Omega_0}{2 \gamma_{\rm eff} t_f}\, \sin\left(\frac{\pi t}{t_f}\right), \quad 0 \le t \le t_f.
    \label{eq:cd_pulse_profile}
\end{equation}
Since $\dot{\Omega}(0) = \dot{\Omega}(t_f) = 0$, the auxiliary pulse vanishes at the temporal boundaries, isolating the battery at the end of the protocol. The resulting pulse profiles are shown in Fig.~\ref{fig:pulse_profiles}.
%%%%%%%%%%%%%%%%%%%%%%%%%%%%
%بررسی دینامیک شارژ برای سه حالت

\begin{figure}[t!]
    \centering
    \includegraphics[width=\linewidth]{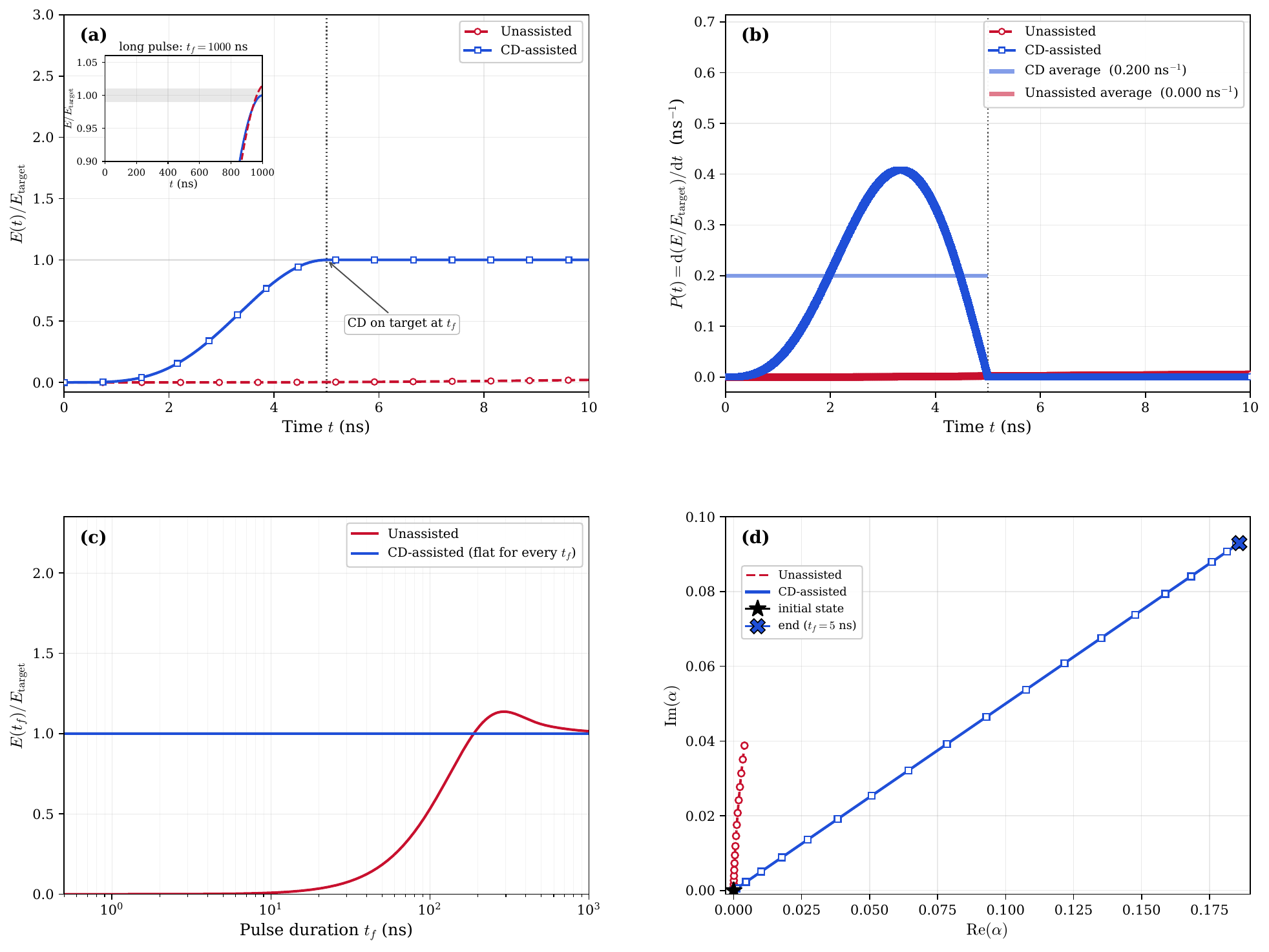}
   \caption{\textbf{CD-assisted versus unassisted charging of the cavity battery } 
(a) Stored energy $E(t)/E_{\rm target}$ for $t_f = 5$ ns (main panel) and $t_f = 1000$ ns (inset).  (b) Charging power $P(t)$ for $t_f = 5$ ns. Horizontal bars show the average power over the ramp, $E(t_f)/(E_{\rm target}t_f)$, for each protocol.  (c) Final energy $E(t_f)/E_{\rm target}$ versus $t_f$ (log scale).  (d) Phase-space trajectory of $\alpha(t)$ over $0 \le t \le 5$ ns.  The fixed physical parameters   are given in Table~\ref{tab:parameters}
}
    \label{fig:dynamic_general_charge}
\end{figure}

%To evaluate the effect on energy storage, we assess the charging performance and coherence maintenance by comparing three driving scenarios: the \textit{adiabatic reference}, defined by $\alpha_{\rm ad}(t) = i\lambda\Omega(t)/\gamma_{\rm eff}$, which reaches the target population without transient excitations; the \textit{unassisted protocol}, where the smooth envelope $\Omega(t)$ is applied directly; and the \textit{CD-assisted protocol}, employing the modified field $\Omega_{\rm CD}(t)$ to enforce exact tracking along $\alpha_{\rm ad}(t)$.
%00000000000000
Fig.~\ref{fig:dynamic_general_charge} compares the charging dynamics under CD-assisted and unassisted driving. Panel (a) shows the stored energy normalized to the target energy, panel (b) presents the normalized instantaneous charging power, panel (c) shows the final stored energy as a function of pulse duration, and panel (d) illustrates the corresponding phase-space trajectories.

For a short pulse, $t_f=5\ \mathrm{ns}$, Fig.~\ref{fig:dynamic_general_charge}~(a) shows that the unassisted protocol is far from the adiabatic regime and that the cavity responds only weakly to the drive, and the stored energy remains below $2\%$ of $E_{\rm target}$ even at $t=10\ \mathrm{ns}$. In contrast, the CD protocol raises the energy monotonically and without oscillations, reaching $E_{\rm target}$ exactly at $t=t_f$ and remaining at the target level thereafter. The inset of (a) shows that for a long pulse, $t_f=1000\ \mathrm{ns}$, the CD and unassisted trajectories nearly coincide and both end close to the target, as expected in the adiabatic limit, where the CD correction, being proportional to $\dot\Omega$, becomes negligible.

Fig.~\ref{fig:dynamic_general_charge}(b) shows the corresponding charging power. For the CD protocol, $P(t)$ is smooth, peaks at $0.41\ \mathrm{ns}^{-1}$ near $t=3.3\ \mathrm{ns}$, and returns to zero at $t_f$, so that charging stops as soon as the target is reached. The horizontal lines indicate the average power over the ramp, $\bar P=E(t_f)/(E_{\rm target}\,t_f)$, which is $0.200\ \mathrm{ns}^{-1}$ for CD (i.e., $1/t_f$) and approximately zero for the unassisted protocol.

The practical advantage of CD driving is its timing independence, shown in Fig.~\ref{fig:dynamic_general_charge}(c). The CD protocol reaches the target energy for every considered $t_f$, whereas the unassisted energy depends strongly on the pulse duration. It is only $\sim\!1\%$ of $E_{\rm target}$ at $t_f=10\ \mathrm{ns}$, reaches the target only near $t_f\approx 2\times10^{2}\ \mathrm{ns}$, overshoots to $\approx1.13\,E_{\rm target}$ at $t_f\approx 3\times10^{2}\ \mathrm{ns}$, and relaxes slowly toward the target (still $\approx 2\%$ above it at $t_f=10^{3}\ \mathrm{ns}$). Hence, unassisted charging requires fine tuning of $t_f$ or very long pulses, while CD charging does not. This robustness comes at the price of an additional control field proportional to $\dot\Omega$, whose cost is analyzed in Sec.~\ref{subsec:cost_power_tradeoff}.

Finally, Fig.~\ref{fig:dynamic_general_charge}(d) shows $\alpha(t)$ in phase space. The CD trajectory follows a nearly straight line from the initial vacuum state to the target coherent state at $\alpha(t_f)\approx 0.19+0.09\,i$, i.e., a direct displacement at nearly constant phase. The unassisted trajectory remains close to the origin ($|\alpha|\lesssim 0.04$) over the same interval, indicating that the battery does not reach the target energy on this timescale.
%As Fig.~(\ref{fig:dynamic_general_charge}) shows for short charging durations ($t_f \sim 1/\kappa$), the unassisted drive exhibits significant transient lag and phase-mismatch oscillations . Conversely, the CD-assisted protocol suppresses non-adiabatic excitations throughout $t \in [0, t_f]$, strictly following the adiabatic trajectory. As the ramp time enters the quasi-adiabatic regime ($t_f \gg 1/\omega_{\rm eff}$), the unassisted trajectory converges toward the target, yet suffers from cumulative dissipation due to cavity decay ($\kappa$). The CD-assisted protocol resolves this trade-off, enabling operation in the ultra-fast regime ($t_f \ll 1/\kappa$) where open-system dissipation is minimized, while retaining the tracking fidelity of slow adiabatic processes.
%

The final stored energy of the cavity depends on the effective detuning
$\omega_{\rm eff}$ through the target-energy scale
$E_{\rm target} = \omega_a \lambda^2 \Omega_0^2 / (\omega^2_{\rm eff}+\kappa^2/4 )$.
Thus, driving closer to the effective cavity resonance  increases the accessible
stored energy for a given field amplitude. However, the dispersive approximation requires $|\Omega(t)/\Delta_d|\ll1$, which bounds the drive amplitude $\Omega_0$ and thereby limits the maximum achievable  $E_{\rm target}$
within the regime of validity. 

The primary objective of this work is not to maximize the absolute stored energy, but to achieve reliable charging to a prescribed target through CD control. The CD protocol reaches this target monotonically over the considered range of pulse durations, whereas unassisted charging is more sensitive to timing and may fall substantially below the target for short pulses.
%==============================
%==============================
%==============================
\subsection{Geodesic Efficiency}
\label{sec:geo}
We quantify the directness of the state evolution using the
Bures  distance \cite{Bures}. The shortest possible path between $|0\rangle$ and $|\alpha_f\rangle$
in Hilbert space has length $L_{\rm geo}=\arccos e^{-|\alpha_f|^2/2}$, the actual evolution follows a path of length $L=\int_0^{t_f}|\dot\alpha|\,dt$. We define
\begin{equation}
  \eta_{\rm geo}=\frac{L_{\rm geo}}{L}
  =\frac{\arccos e^{-|\alpha_f|^2/2}}{\int_0^{t_f}|\dot\alpha|\,dt}\le 1 .
\end{equation}
In our driven–dissipative cavity model, the evolution is restricted to the manifold of coherent states, $\rho(t)=|\alpha(t)\rangle\langle\alpha(t)|$ (Appendix~\ref{app:coherent_state_proof}). Consequently, the length of any accessible trajectory connecting the initial and final states satisfies \(L=\int_0^{t_f}|\dot{\alpha}(t)|dt\geq|\Delta\alpha|\), where \(\Delta\alpha=\alpha_f-\alpha_i\). This inequality establishes an upper bound on the geometric efficiency, \(\eta_{\mathrm{geo}}\leq\eta_{\max}=\arccos(e^{-|\Delta\alpha|^2/2})/|\Delta\alpha|\), consistent with the analysis of Ref. [23]. Under CD driving, the cavity amplitude evolves according to \(\alpha_{\mathrm{CD}}(t)=i\lambda\Omega(t)/\gamma_{\mathrm{eff}}\). Since the reference envelope increases monotonically, the resulting trajectory is a straight line in phase space, with length \(L_{\mathrm{CD}}=|\Delta\alpha|\). The CD protocol therefore saturates the upper bound, \(\eta_{\mathrm{geo}}^{\mathrm{CD}}=\eta_{\max}\), achieving the shortest possible trajectory within the accessible coherent-state manifold, although this constrained geometric optimum need not coincide with the shortest path in the full Hilbert space.
%%%%%%%%%%%%%%%%%%%%%%%%%%%%%
%unrestricted quantum speed-limit bound: حداقل زمان بنیادی لازم برای انتقال بین دو حالت کوانتومی، وقتی هیچ محدودیتی به خانوادهٔ حالت‌ها یا شکل کنترل اعمال نمی‌شود. در این حالت، همهٔ مسیرهای فیزیکی مجاز در فضای هیلبرت، و نه فقط مسیرهای حالت همدوس، در نظر گرفته می‌شوند.

%نکتهٔ اصلی این است که دو نوع بهینگی از هم جدا هستند:

%پروتکل CD ممکن است در فضای محدود حالت‌های همدوس، کوتاه‌ترین یا مستقیم‌ترین مسیر را ایجاد کند.اما این نتیجه به‌تنهایی ثابت نمی‌کند که CD در فضای کامل حالت‌های کوانتومی سریع‌ترین مسیر ممکن را طی می‌کند.به بیان دیگر، CD از نظر هندسی در یک کلاس محدود از مسیرها بهینه است، اما این بهینگی الزاماً به معنای بهینگی مطلق از نظر سرعت کوانتومی نیست. ممکن است پروتکلی که از حالت‌های غیرهمدوس، برهم‌نهی‌های پیچیده‌تر یا مسیرهایی خارج از منیفلد حالت‌های همدوس استفاده می‌کند، در اصل انتقال را در زمان کوتاه‌تری انجام دهد.
%%%%%%%%%%%%%%%%%%%%%%%%%%5
\begin{figure}[t!]
    \centering
    \includegraphics[width=\linewidth]{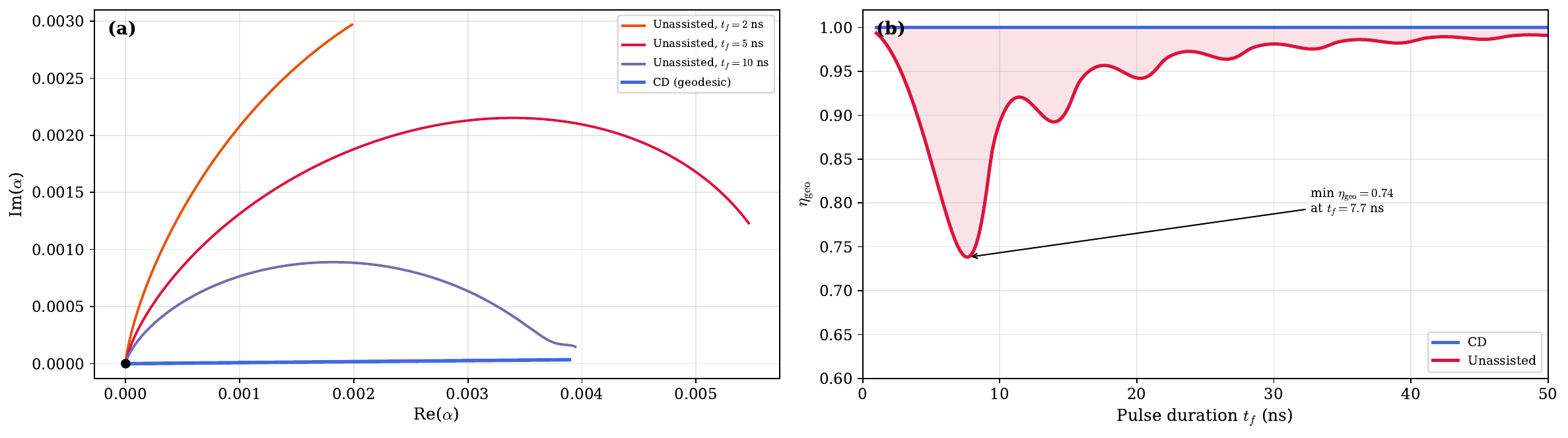}
	\caption{\textbf{Geodesic efficiency of CD and unassisted charging.}
(a)~Phase-space trajectories of the cavity amplitude $\alpha$. The CD trajectory (blue) follows a straight line from the origin to the target state. The unassisted paths (colored) spiral because of the cavity lag, which adds path length for $t_f=2$, 5 and 10~ns.
(b)~Geodesic efficiency $\eta_{\rm geo}$ against pulse duration $t_f$. CD stays at $\eta_{\rm geo}\simeq0.9998$ for all $t_f$. The unassisted protocol drops to a minimum of $\approx0.74$ near $t_f\simeq 7.7$~ns and recovers for slow pulses. The shaded region highlights  the difference between the two protocols.}
    \label{fig4new:geodesic_efficiency}
\end{figure}
%=========================================================================
%=========================================================================
%=========================================================================
\subsection{Speed–Cost Trade-off and Optimal Charging Duration}\label{subsec:cost_power_tradeoff}

Implementing the auxiliary CD drive requires additional control resources. We quantify this  cost by  the time-integrated intensity of the driving field~\cite{Werschnik},
%We  adopt a hardware-grounded metric defined by the integrated control field fluence as 
\begin{equation}
    \mathcal{C} = \int_{0}^{t_f} |\Omega(t)|^2\,dt.
    \label{eq:total_fluence_cost}
\end{equation}
We define the control fluence as Eq.~\eqref{eq:total_fluence_cost} in units of ${\mathrm rad}^2/ns$.  It is proportional to the
microwave energy delivered to the cavity only after a line-specific
calibration. No such calibration is performed here. We therefore use $\mathcal{C}$ only to compare protocols
on the same drive line.

When the CD field is applied through a separate control channel, the two drives
contribute independently to the total microwave power, so the total control
fluence decomposes as into the reference contribution $\mathcal{C}_{\mathrm{ref}}$ and the auxiliary CD contribution $\mathcal{C}_{\mathrm{CD}}$
\begin{equation}
    \mathcal{C} = \mathcal{C}_{\mathrm{ref}} + \mathcal{C}_{\mathrm{CD}} = \int_{0}^{t_f} |\Omega_{\mathrm{ref}}(t)|^2\,dt + \int_{0}^{t_f} |\Omega_{\mathrm{CD}}(t)|^2\,dt.
\end{equation}
In a circuit-QED device this condition is satisfied when the reference
and CD tones are delivered through distinct input ports.

For the $\sin^2$-shape reference pulse in Eq.(\ref{eq:adiabatic_pulse_def}),  analytical integration  yields
\begin{equation}
    \mathcal{C}_{\mathrm{ref}} = \frac{3}{8}\Omega_0^2\,t_f, \hspace{1cm} \mathcal{C}_{\mathrm{CD}} = \frac{\pi^2 \Omega_0^2}{8 |\gamma_{\mathrm{eff}}|^2\,t_f},
    \label{eq:cost_ref_exact}
\end{equation}
for the reference and auxiliary CD fields, respectively.
%Dimensionally, $[\mathcal{C}] = \mathrm{time}^{-1}$ (expressed in $\mathrm{ns}^{-1}$), which directly maps to energy via the physical control line calibration $E_{\mathrm{cost}} = \xi \hbar \mathcal{C}$, where $\xi$ is a dimensionless hardware factor. 
The opposing scalings $\mathcal{C}_{\rm ref}\propto t_f$ and
$\mathcal{C}_{\rm CD}\propto 1/t_f$  establish a speed-cost trade-off in the charging protocol,
qualitatively consistent with \cite{campbell_2017}, although the cost measure differs.  
%از این رابطه عملا استفاده نشده و مقدار عددی آن هم به دست نیامده است. بنابراین کلا حذفش می کنیم
%Combining this with the stored ergotropy defines the overall operational charging efficiency as
%\begin{equation}
 %   \eta_{\mathrm{ch}} \equiv \frac{E_{\mathrm{stored}}(t_f)}{E_{\mathrm{cost}}(t_f)} = \frac{\hbar \omega_a |\alpha(t_f)|^2}{\eta \hbar (\mathcal{C}_{\mathrm{ref}} + \mathcal{C}_{\mathrm{CD}})}.
 %   \label{eq:charging_efficiency}
%\end{equation}

In the ultrafast regime ($t_f \to 0$), the auxiliary drive cost diverges, requiring unphysically large microwave amplitudes and substantially diminishing the charging efficiency. Conversely, in the long-time adiabatic limit ($t_f \gg \kappa^{-1}$), although the CD overhead vanishes ($\mathcal{C}_{\mathrm{CD}} \to 0$), cavity dissipation at rate $\kappa$ degrades the stored energy while the reference cost grows linearly with duration. Balancing these competing mechanisms, minimizing the total field cost yields a well-defined optimal charging duration
\begin{equation} \label{t_opt}
t_f^{\mathrm{opt}} = \frac{\pi}{\sqrt{3}\,|\gamma_{\mathrm{eff}}|}
\end{equation}
which gives $ \mathcal{C}_{\rm tot}^{\min} =\frac{\sqrt{3}\pi}{4}\frac{\Omega_0^{2}}{|\gamma_{\rm eff}|}$. 
%%%%%اضافه شد
The optimal charging duration in Eq.~\eqref{t_opt} depends on both the effective cavity detuning and the cavity decay rate through \(|\gamma_{\mathrm{eff}}|=\sqrt{\omega_{\mathrm{eff}}^2+\kappa^2/4}\). 
 In the detuning-dominated
limit, $\kappa/|\omega_{\mathrm{eff}}|\ll 1$, the optimal time approaches
$\pi/(\sqrt{3}\,|\omega_{\mathrm{eff}}|)$. In the resonant limit,
$\omega_{\mathrm{eff}}\rightarrow 0$, it approaches
$2\pi/(\sqrt{3}\,\kappa)$. 
 %In the strongly detuned regime, \(|\omega_{\mathrm{eff}}|\gg\kappa\), the optimal duration is primarily determined by the coherent cavity dynamics, and the influence of dissipation is comparatively weak. However, this condition is not required by the qubit–cavity dispersive approximation. By tuning the driving frequency close to the dispersively shifted cavity resonance while maintaining \(g\ll|\Delta|\), one can access a regime in which \(\kappa\) becomes comparable to \(|\omega_{\mathrm{eff}}|\). In this regime, cavity damping contributes substantially to the dynamical response and modifies the optimal charging duration. 
%To characterize the energetic consequences of dissipation, we evaluate the cumulative cavity energy loss during ideal counterdiabatic charging. For the prescribed \(\sin^2\) reference envelope, the exact CD trajectory gives \(W_{\mathrm{loss}}=(3/8)\kappa E_f t_f\), yielding an effective energy-transfer efficiency \(\eta_{\mathrm{transfer}}=[1+3\kappa t_f/8]^{-1}\). These expressions distinguish the minimum-fluence operating time from the efficiency of energy transfer to the battery and provide a quantitative measure of the influence of dissipation on the charging process.
%%%%%اضافه شد
%%%%%%%%%%%%%%%%%%

Fig.~\ref{fig:dissipation_analysis}(a) shows the optimal pulse duration, $t_f^{\mathrm{opt}}$, increases with the damping-to-detuning ratio $\kappa/\omega_{\mathrm{eff}}$.  

As shown in Fig.~\ref{fig:dissipation_analysis}(b), the total field fluence
$C_{\mathrm{tot}}=C_{\mathrm{ref}}+C_{\mathrm{CD}}$ reaches a minimum at
the corresponding optimal duration in each regime. The markers identify these
analytical minima for the weak-damping and near-resonant parameter choices.
This behavior reflects the competition between the reference-pulse
contribution, which increases with duration, and the CD contribution, which
increases as the protocol duration is reduced.

The effect of dissipation on energy transfer is quantified in
Fig.~\ref{fig:dissipation_analysis}(c). We define the transfer efficiency
as the ratio of the final stored energy to the total energy supplied,
$\eta_{\mathrm{transfer}}=E_f/(E_f+E_{\mathrm{diss}})$. Within the
adiabatic effective-cavity model, the $\sin^2$ pulse envelope gives
\begin{equation}
    \eta_{\mathrm{transfer}}
    = \frac{1}{1+3\kappa t_f/8},
    \qquad
    \frac{E_{\mathrm{diss}}}{E_f}
    = \frac{3\kappa t_f}{8}.
\end{equation}
Thus, the transfer efficiency decreases as the dissipative exposure
$\kappa t_f$ increases. The marked operating points show the transfer efficiencies at the respective fluence-optimal durations, illustrating the trade-off between control-cost minimization and energy dissipation. 

\begin{figure}[t]
    \centering
    \includegraphics[width=1\linewidth]{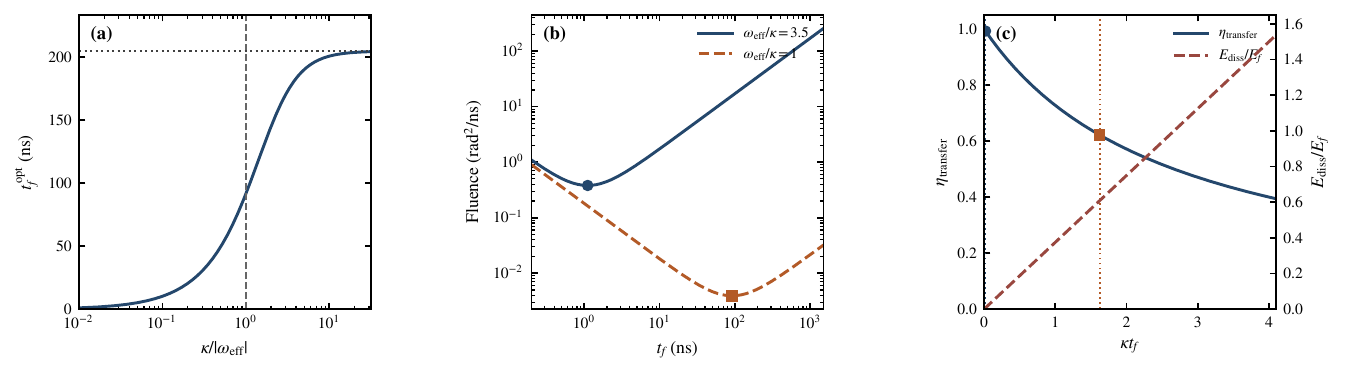}
    \caption{\textbf{
  Control cost and dissipation-dependent performance  of CD charging. }
    (a) Optimal charging time,
    $t_f^{\mathrm{opt}}=\pi/(\sqrt{3}\,|\gamma_{\mathrm{eff}}|)$,
    as a function of the damping-to-detuning ratio
    $\kappa/|\omega_{\mathrm{eff}}|$.
    The vertical dashed line marks
    $\kappa/|\omega_{\mathrm{eff}}|=1$; the horizontal dotted line indicates
    the resonant limit ($\omega_{\mathrm{eff}}=0$) $t_f^{\mathrm{opt}}=2\pi/(\sqrt{3}\,\kappa)$.
    (b) Total field fluence,
    $C_{\mathrm{tot}}=C_{\mathrm{ref}}+C_{\mathrm{CD}}$,
    as a function of the charging duration $t_f$.% for a fixed target amplitude$|\alpha_f|=0.005$. 
The curves correspond to the weak-damping  (blue)
    ($\omega_d/2\pi=5.000\,\mathrm{GHz}$) and the $\omega_{\mathrm{eff}}/\kappa \approx1$  (red) regimes.
    Markers indicate the analytically determined optimal durations and their corresponding minimum fluences.
    (c) Energy-transfer efficiency
    $\eta_{\mathrm{transfer}}=(1+3\kappa t_f/8)^{-1}$ (left axis) and
    normalized dissipated energy
    $E_{\mathrm{diss}}/E_f=3\kappa t_f/8$ (right axis), plotted against the
    dissipative exposure $\kappa t_f$. The vertical dotted lines and markers
    indicate the optimal operating points.
    }
    \label{fig:dissipation_analysis}
\end{figure}

Beyond the duration dependence of the control fluence, the interplay between CD control effort and photon leakage imposes an additional constraint on the charging protocol. The  energy dissipated during charging is \(E_{\mathrm{diss}}=\kappa\omega_a\int_0^{t_f}|\alpha(t)|^2dt\). Combining this expression with the geometric CD cost, \(C_{\mathrm{CD}}=\lambda^{-2}\int_0^{t_f}|\dot{\alpha}(t)|^2dt\), and using the Cauchy-Schwarz inequality together with the vacuum initial condition, we obtain the lower bound (see Appendix~\ref{Lower bound} for the derivation)
\[
C_{\mathrm{CD}}E_{\mathrm{diss}}
\geq
\frac{\kappa E_f^2}
{4\omega_a\lambda^2},
\]
where \(E_f=\omega_a|\alpha_f|^2\) is the final stored energy. This bound establishes that, for a fixed target energy, the control effort and dissipative losses cannot be simultaneously minimized without limit.
In particular, suppressing dissipation by reshaping the charging envelope so that 
$\alpha(t)$
 remains small for most of the interval and rises sharply near 
$t_f$ generally requires a larger  $|\dot{\alpha}(t)|$,  which increases the CD control cost. This analytical constraint complements the optimal-duration analysis by revealing an intrinsic trade-off between control effort and dissipation, irrespective of the specific charging envelope.
%=========================================================================
%=========================================================================
%=========================================================================

%==================================================
\section{Self-Discharging Dynamics and Ergotropy Retention}

Upon reaching the target charging time $t_f$, the external driving field is switched off ($\Omega(t) = 0$ for $t \ge t_f$), and the qubit-cavity dispersive coupling is deactivated ($g \to 0$), which can be experimentally achieved by detuning the qubit or employing tunable couplers. Alternatively, satisfying the condition $2\omega_q = \omega_a + \omega_d$ effectively decouples the charging mechanism ($\lambda = 0$). Consequently, the cavity battery enters the self-discharge phase, governed by the effective Hamiltonian $H_{\rm eff} = \omega_{\rm eff} a^\dagger a$. 

In the absence of coherent driving, the uncoupled cavity mode interacting with a zero-temperature reservoir is described by the Lindblad master equation as
\begin{equation}
    \frac{d\rho_a}{dt} = -i[\omega_{\rm eff} a^\dagger a, \rho_a] + \kappa \mathcal{D}[a]\rho_a,
    \label{eq:cavity_decay_lindblad}
\end{equation}
The expectation value of the annihilation operator, $\alpha(t) \equiv \langle a(t) \rangle$, satisfies the following autonomous first-order equation of motion
\begin{equation}
    \dot{\alpha}(t) = -\left( i\omega_{\rm eff} + \frac{\kappa}{2} \right) \alpha(t), \quad (t \ge t_f).
    \label{eq:alpha_decay_ode}
\end{equation}
Integrating Eq.~\eqref{eq:alpha_decay_ode} yields the exact analytical trajectory for the complex field amplitude:
\begin{equation}
    \alpha(t) = \alpha(t_f) \exp\left[-\left( i\omega_{\rm eff} + \frac{\kappa}{2}\right)(t - t_f)\right],
    \label{eq:alpha_decay_sol}
\end{equation}
where $\alpha(t_f)$ is the amplitude achieved at the conclusion of the charging protocol.

From Eq.~\eqref{eq:alpha_decay_sol}, it is evident that the field amplitude and the stored energy exhibit distinct relaxation timescales:
\begin{align}
    |\alpha(t)| &= |\alpha(t_f)|\, e^{-(t-t_f)/\tau_\alpha}, \quad \tau_\alpha = \frac{2}{\kappa}, \label{eq:tau_alpha} \\
    n(t) &= |\alpha(t)|^2 = n(t_f)\, e^{-(t-t_f)/\tau_E}, \quad \tau_E = \frac{1}{\kappa}. \label{eq:tau_energy}
\end{align}
While the field amplitude relaxes with the characteristic time $\tau_\alpha = 2/\kappa$, the mean photon number $n(t)$ and the stored energy $E(t) = \hbar\omega_a n(t)$ deplete at twice that rate, establishing the fundamental energy lifetime of the battery as $\tau_E = \kappa^{-1}$.
%**************************************
A pivotal thermodynamic question during the self-discharge stage is whether the dissipation of energy compromises the extractable work. For an arbitrary battery state $\rho_a$, the maximum extractable work via cyclic unitary operations is quantified by the ergotropy:
\begin{equation}
    \mathcal{E}(\rho_a) = \operatorname{Tr}[H_a \rho_a] - \min_{U \in \mathcal{U}} \operatorname{Tr}[H_a U \rho_a U^\dagger],
    \label{eq:ergotropy_def}
\end{equation}
where $H_a = \hbar\omega_a a^\dagger a$. Because the zero-temperature dissipator $\mathcal{D}[a]$ maps coherent states into coherent states, the cavity remains strictly in a pure coherent state $|\alpha(t)\rangle$ throughout the uncoupled discharge stage:
\begin{equation}
    \rho_a(t) = |\alpha(t)\rangle \langle \alpha(t)| \implies \operatorname{Tr}[\rho_a(t)^2] = 1, \quad \forall t \ge t_f.
    \label{eq:pure_coherent_state}
\end{equation}
Since any pure state can be unitarily mapped to the vacuum state $|0\rangle$ via the displacement operator $D(-\alpha(t))$, the passive state associated with $\rho_a(t)$ is uniquely the vacuum state $|0\rangle\langle 0|$, which carries zero energy. Consequently, all stored energy in the cavity battery remains entirely extractable as coherent work, leading to the exact identity:
\begin{equation}
    \mathcal{E}(t) = E(t) = \hbar\omega_a |\alpha(t_f)|^2 e^{-\kappa(t - t_f)}.
    \label{eq:ergotropy_equality}
\end{equation}

%fig

\begin{figure}[htb!]
\centering
\includegraphics[width=\linewidth]{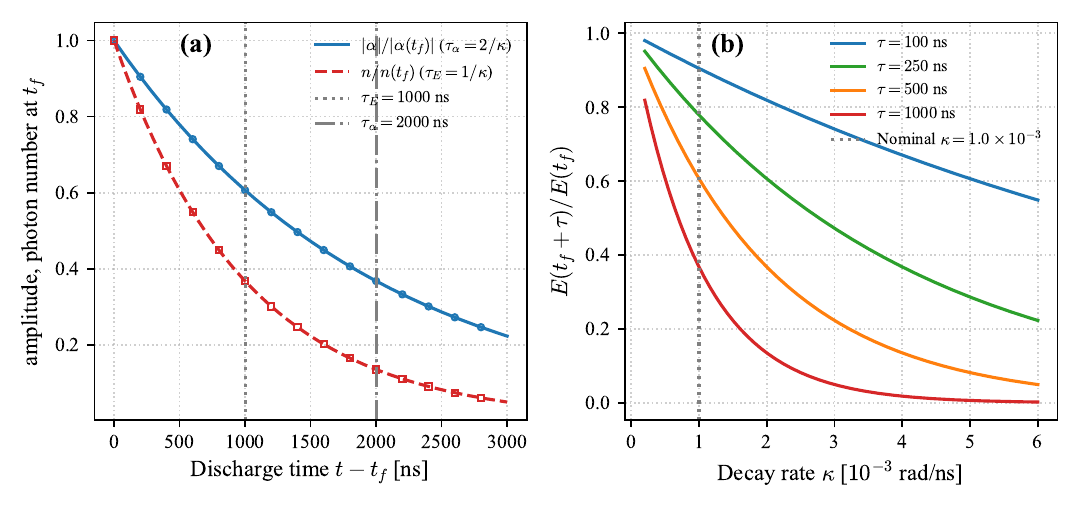}
\caption{Spontaneous self-discharge dynamics and  energy retention of the cavity quantum battery.
(a) Time evolution of the normalized coherent amplitude $|\alpha|/|\alpha(t_f)|$  and photon population $n/n(t_f)$ after charging, illustrating their characteristic decay timescales $\tau_\alpha = 2/\kappa$ and  $\tau_E = 1/\kappa$, respectively. (b) Normalized retained energy \(E(t_f+\tau)/E(t_f)\) as a function of the cavity decay rate \(\kappa\) for different storage durations \(\tau\). }
\label{fig:self_discharge}
\end{figure}
%**************************************
%caption of the fig
The comprehensive operational cycle of the battery, encompassing both driven charging and subsequent spontaneous discharging, is depicted in Fig.~\ref{fig:self_discharge}. 
Upon reaching the target charging time $t_f$, the charging is switched off, and the system enters an unperturbed, open-storage regime governed by the Lindblad dissipator~\eqref{eq:cavity_decay_lindblad}.

%Figure~\ref{fig:self_discharge}(a) captures the uninterrupted trajectory in phase space. The active counterdiabatic protocol smoothly steers the cavity field from the vacuum state $|0\rangle$ to the target displacement $\alpha(t_f)$. Once the drive is extinguished, the field exhibits an inward spiral towards the origin. %This spiral morphology originates from the persistent effective frequency shift $\omega_{\mathrm{eff}} = \omega_a' - g^2/\Delta_a$, which imparts a continuous coherent phase rotation on top of the dissipative shrinkage of the radius.

A fundamental aspect of this bosonic open battery is the clear disparity between observable lifetimes, as quantified in Fig.~\ref{fig:self_discharge}(a). 
While the coherent field amplitude $|\alpha(t)|$ decays according to the characteristic timescale $\tau_\alpha = 2/\kappa$ 
%($\tau_\alpha = 2000\,\mathrm{ns}$ for nominal parameters)
, the photon population $n(t) = |\alpha(t)|^2$ and consequently the stored energy $E(t) = \hbar\omega_a n(t)$ decay at twice that rate, with a lifetime $\tau_E = 1/\kappa = 1000\,\mathrm{ns}$. 
Since the open dynamics at zero temperature strictly preserves coherent state purity ($\operatorname{Tr}[\rho_c^2] = 1$), the extractable ergotropy remains identical to the total stored energy, $\mathcal{E}(t) = E(t)$, confirming that the operational storage lifetime of extractable work is determined solely by $\tau_E$.

To assess battery resilience against cavity losses during storage, Fig.~\ref{fig:self_discharge}(b) presents the  sensitivity as a function of the decay rate $\kappa$ for multiple idle holding durations $\tau = t - t_f$. 
For our baseline parameter $\kappa  \sim 1.0 \times 10^{-3}\,\mathrm{rad/ns}$, the battery retains approximately $90.5\%$ of its energy after an idle time of $\tau = 100\,\mathrm{ns}$, and $36.8\%$ ($1/e$) at $\tau = \tau_E = 1000\,\mathrm{ns}$. This demonstrates that high-Q superconducting resonators provide a sufficiently broad storage window for subsequent on-demand work extraction.

%Finally, Fig.~\ref{fig:self_discharge}(d) contrasts the complete charge-discharge trajectories under the CD and standard protocols. 
%Although the CD driving provides a substantial boost in the peak energy $E(t_f)$ achieved within the short deadline $t_f = 50\,\mathrm{ns}$, the post-shutoff decay exponent is identical for both schemes. This establishes that shortcut-to-adiabaticity protocols enhance charging power and energetic capacity without modifying the autonomous retention capability. %, which is strictly governed by the passive quality factor $Q = \omega_a/\kappa$ of the cavity.

%**************************************

\section{Experimental Feasibility and Physical Parameter Regimes.}
\label{sec_experimental}
Although our primary model is formulated in the dispersive regime of circuit quantum electrodynamics (cQED), the underlying charging dynamics and the proposed CD  protocols can also be mapped onto other  experimental platforms, including trapped-ion motional modes and superconducting quantum-acoustic resonators \cite{Zhou2017}. In a trapped-ion implementation, the quantized vibrational motion acts as the battery,the ion's quantized vibrational motion serves as the battery while a phase-controlled optical force generates the required coherent displacement. In an appropriate rotating frame, the effective  Hamiltonian takes the form \(H=\omega b^\dagger b-[F(t)b^\dagger+F^*(t)b]\), where \(b\) denotes the phonon annihilation operator and \(F(t)\) is the controllable driving force.  An analogous CD correction can be constructed for this platform.  CD  driving of a single $^{171}\mathrm{Yb}^+$ ion has been experimentally demonstrated to generate fast phase-space displacements \cite{An2016nature}.

Furthermore, quantized mechanical motion has been investigated experimentally as a quantum-battery platform, supporting its suitability for investigating
coherent energy storage and ergotropy \cite{Wang2024}. The framework can also be extended to superconducting quantum-acoustic architectures, where a
transmon qubit couples to a mechanical resonator to mediate the charging
process \cite{Manenti2017}. While these platforms offer promising opportunities to test the generality of the proposed charging protocol, a quantitative assessment of experimental
performance will require accounting for realistic motional heating, finite
control bandwidth, and auxiliary mediator dynamics.

%=====================================================
\section{Conclusion and discussion}
\label{sec_conclusion}

We have investigated the charging dynamics of a switchable cavity quantum battery coupled dispersively to a driven qubit. At zero temperature, the cavity remains in a pure coherent state during both charging and self-discharge. Consequently, the stored energy coincides with the extractable ergotropy throughout the evolution. Under constant driving, the final stored energy is highly sensitive to the charging duration, such that small timing deviations can lead to substantial variations in the stored energy, ranging from nearly zero to its maximum value.

 A \(\sin^2\)-shaped adiabatic charging pulse enables the cavity to follow the instantaneous steady-state trajectory in phase space. However, achieving sufficiently accurate adiabatic following requires relatively long charging durations, during which cavity dissipation reduces the stored energy. To address this limitation, we introduced a CD protocol that enables exact tracking of the prescribed trajectory within the effective model. In contrast to unassisted charging, for which the final stored energy depends strongly on the pulse duration, the CD protocol reaches the desired target state independently of the charging duration, provided the required control field can be implemented. This robustness requires  an additional control field proportional to the time derivative of the reference drive.

In the ultrafast regime (\(t_f\rightarrow0\)), the auxiliary-field amplitude increases significantly, leading to a divergent control cost in the idealized model and rendering such protocols impractical under finite-amplitude constraints. Conversely, in the long-time adiabatic limit (\(t_f\gg\kappa^{-1}\)), the counterdiabatic overhead vanishes, whereas cavity dissipation progressively degrades the stored energy and the reference-field cost increases with the charging duration. The competition between these contributions yields an optimal charging duration that minimizes the total control cost within the considered cost model.

The geometric analysis provides further insight into the CD protocol. In the drive-rotating frame, the CD trajectory follows a straight, non-backtracking path in the coherent-state manifold, thereby saturating the corresponding geometric-efficiency bound. By contrast, unassisted charging generally involves additional phase-space motion arising from the finite response time of the cavity. This geometric characterization provides insight into the trajectory optimization achieved by CD driving, although geometric optimality alone does not establish optimality with respect to charging time, control cost, or energy storage.

After charging, the stored energy decreases because of cavity dissipation, with the cavity decay rate determining the energy-retention timescale and thus the available storage window for subsequent energy extraction. The accessible stored energy can be enhanced by tuning the driving frequency closer to the effective cavity resonance, provided that the dispersive approximation remains valid and the required control fields remain experimentally feasible.
%اضافه شد باند
Furthermore, the derived lower bound on the product of counterdiabatic control cost and dissipative losses highlights the unavoidable trade-off between reducing energy dissipation and limiting control effort for a fixed charging target, complementing the optimal charging-duration analysis.
%اضافه شد باند
Finally, the analysis is based on an effective cavity model with idealized control assumptions. Future studies should incorporate multilevel transmon dynamics, finite control bandwidth, realistic dissipation, and experimentally constrained control amplitudes to assess the practical feasibility of the proposed protocol.

\section{Appendix}
\subsection{Derivation of the Effective Hamiltonian}
\label{app:derivation_effective}

In this appendix, we derive the effective Hamiltonian in Eq.~\ref{eq:Heff_final}  from Eq.~\ref{eq:rotating_H}.

We introduce a unitary transformation $\hat{U}=e^ {\hat{S}} $ satisfying
\begin {equation} \label{equation1}
 [\hat{S}, \hat{H}_0] = \hat{H}_I
\end{equation} 
where $\hat{H}_I = \hbar g (\hat{a}^\dagger \hat{\sigma}^- + \hat{a} \hat{\sigma}^+) + \Omega(t) (\hat{\sigma}^- + \hat{\sigma}^+)$.
We choose
$
\hat{S} = \alpha (\hat{a}^\dagger \hat{\sigma}^- - \hat{a} \hat{\sigma}^+) + \beta (\hat{\sigma}^- - \hat{\sigma}^+)
$.
Using Eq.~\ref{equation1} we obtain
\begin{equation}
\alpha = \frac{g}{\Delta}, \qquad \beta = \frac{\Omega(t)}{\Delta_q}.
\end{equation}
Applying  the Frohlich-Nakajima transformation to Eq.~\ref{eq:rotating_H} yields
\begin{equation}
\hat{H}_{\rm eff} =\hat{U}^{\dag}\hat{H}\hat{U}= \hat{H}_0 + \frac{1}{2}[\hat{H}_I, \hat{S}].
\end{equation}
to Eq.~\ref{eq:rotating_H} and get 

\begin{align}
\hat{H}_{\rm eff} &= \frac{1}{2} \Delta_q \sigma_z
+  \Delta_a \hat{a}^\dagger \hat{a}
+ \frac{g^2}{\Delta} (\hat{\sigma}^+ \hat{\sigma}^- + \hat{a}^\dagger \hat{a} \sigma_z) \nonumber \\
&+ \frac{\Omega^2(t)}{\Delta_q} \sigma_z
+ g\Omega(t) \left( \frac{1}{\Delta_q} + \frac{1}{\Delta} \right) (\hat{a}^\dagger + \hat{a}) \sigma_z.
\end{align}
For the qubit ground state, we set \(\sigma_z = -1\) and \(\hat{\sigma}^+ \hat{\sigma}^- = constant\). After neglecting the constant energy shift, we obtain
\begin{equation}
\hat{H}_{\rm eff} = \omega_{\rm eff} \hat{a}^\dagger \hat{a} - \lambda \Omega(t) (\hat{a}^\dagger + \hat{a}),
\end{equation}
where
\begin{equation}
\omega_{\rm eff} = \Delta_a - \frac{g^2}{\Delta}, \qquad
\lambda = g \left( \frac{1}{\Delta_q} + \frac{1}{\Delta} \right).
\end{equation}
This completes the derivation.
%=========================================
\subsection{Coherent State Preservation Under Markovian Decay}
\label{app:coherent_state_proof}

To demonstrate that the battery remains in a pure coherent state under Eq.~\eqref{eq:master_eq}, we transform the density matrix into the displaced frame via the Weyl operator $\hat{D}(\alpha(t)) = \exp[\alpha(t)\hat{a}^\dagger - \alpha^*(t)\hat{a}]$, defining $\hat{\rho}_D(t) \equiv \hat{D}^\dagger(\alpha(t))\hat{\rho}(t)\hat{D}(\alpha(t))$. 

When the complex displacement amplitude $\alpha(t)$ satisfies Eq.~\eqref{eq:eom_alpha}, all linear drive and displacement terms cancel. Consequently, the master equation in the displaced frame reduces to that of an unforced decaying harmonic oscillator:
\begin{equation}
    \frac{d\hat{\rho}_D(t)}{dt} = -i\omega_{\rm eff} [\hat{a}^\dagger\hat{a}, \hat{\rho}_D(t)] + \kappa \mathcal{D}[\hat{a}]\hat{\rho}_D(t).
\end{equation}
For an initially uncharged battery with $\hat{\rho}(0) = |0\rangle\langle 0|$ ($\alpha(0)=0$), the vacuum state is the unique stationary state ($d\hat{\rho}_D/dt = 0$), ensuring $\hat{\rho}_D(t) = |0\rangle\langle 0|$ for all $t \ge 0$. Transforming back to the original frame gives $\hat{\rho}(t) = \hat{D}(\alpha(t)) |0\rangle\langle 0| \hat{D}^\dagger(\alpha(t)) = |\alpha(t)\rangle\langle\alpha(t)|$, confirming unit purity $\text{Tr}[\hat{\rho}^2(t)] = 1$ during charging.
%++++++++++++++++++++++++++++++++++++++++++++++++++
\subsection{Energy and Power}
\label{app:power_derivation}
The instantaneous charging power $P(t)$  is defined as the time derivative of the stored internal energy $E(t) = \omega_a \alpha(t)\alpha^*(t)$,
\begin{equation}
\begin{aligned}
P(t) &= \frac{dE(t)}{dt} = \omega_a \frac{d}{dt} \left( \alpha(t)\alpha^*(t) \right) \\
&= \omega_a \left( \dot{\alpha}(t)\alpha^*(t) + \alpha(t)\dot{\alpha}^*(t) \right).
\end{aligned}
\end{equation}
Substituting the equations of motion for the cavity amplitude and its complex conjugate, we obtain 
\begin{equation}
\begin{aligned}
P(t) &= \omega_a \Big\{ \left( -\gamma_{\text{eff}}\alpha(t) + i\lambda\Omega(t) \right)\alpha^*(t) \\
&\quad + \alpha(t)\left( -\gamma_{\text{eff}}^*\alpha^*(t) - i\lambda\Omega(t) \right) \Big\}
%&= \omega_a \Big\{ -(\gamma_{\text{eff}} + \gamma_{\text{eff}}^*)\vert\alpha(t)\vert^2 + i\lambda\Omega(t)\left(\alpha^*(t) - \alpha(t)\right) \Big\}.
\end{aligned}
\end{equation}
Using $\gamma_{\text{eff}}+\gamma_{\text{eff}}^* =\kappa$ and  $\alpha(t) - \alpha^*(t) = 2i\operatorname{Im}[\alpha(t)]$,the power expression can be separated into drive-induced and dissipative contributions
\begin{equation}
\begin{aligned}
P(t) &= -\omega_a \kappa \vert\alpha(t)\vert^2 + 2\omega_a \lambda \Omega(t)\operatorname{Im}[\alpha(t)] \\
&=P_{\text{in}}(t) - P_{\text{diss}}(t),
\end{aligned}
\end{equation}
where $P_{\text{in}}(t) = 2\omega_a \lambda \Omega(t)\operatorname{Im}[\alpha(t)]$ denotes the power supplied by the drive, and $P_{\text{diss}}(t) = \kappa E(t)$ represents the cavity's dissipative power loss.

%+++++++++++++++++++++++++++++++++++++
\subsection{Analytical Solution of the Equation of Motion}
Direct evaluation of the integral in Eq.~(\ref{eq:alpha_exact_solution}) gives

\begin{align*}\label{eq:alpha_exact_analytic}
\alpha(t) &= \frac{i\lambda \Omega_0}{2} \left[ \frac{1}{\gamma_{\mathrm{eff}}} \left(1 - e^{-\gamma_{\mathrm{eff}}t}\right) \right. \nonumber \\
&\quad \left. -\, \frac{1}{\gamma_{\mathrm{eff}}^2 + \omega_0^2} \left( \gamma_{\mathrm{eff}} \cos(\omega_0 t)  
+ \omega_0 \sin(\omega_0 t) - \gamma_{\mathrm{eff}} e^{-\gamma_{\mathrm{eff}}t} \right) \right]
\end{align*}
where $\omega_0 =\pi/t_f$.
%--------------------------------------------------
%%%%%%%%%%%%%%%%%%%%%%%%%%%%%%%%%%%%%%%%

\subsection{Geodesic Efficiency of the CD Protocol}
\label{app:geo}
%\subsection*{ Mean-field dynamics and the effective unitary}

As shown in Sec.\ref{app:coherent_state_proof}, the cavity state remains coherent state $|\alpha(t)\rangle=D(\alpha(t))|0\rangle$, where
$D(\alpha)=\exp(\alpha a^\dagger-\alpha^* a)$. 
Thus, the evolution of this state is
 described by a unitary operator, $U(t)=D(\alpha(t))$, and the dynamics can be equivalently described by the Hamitonian 
\begin{equation}
  H_{U}=i\dot U U^\dagger
  = i\big(\dot\alpha\,a^\dagger-\dot\alpha^{*}a\big).
  \label{eq:S_H_u}
\end{equation}
 The infinitesimal distance element traversed along the physical trajectory during an interval $dt$ for pure state $|\psi(t)\rangle$ is determined by the Bures metric %reference added ye ketab
\begin{equation}
 ds^2=2(1-|\langle\psi(t)|\psi(t+dt)\rangle|).
  \label{eq:Buress_metric}
\end{equation}
Using Schr\"odinger equation $i|\dot{\psi(t)}\rangle=H_{U}|\psi(t)\rangle$ and $|\psi(t+dt)\rangle=|\psi(t)\rangle+|\dot{\psi(t)}\rangle dt + ...$, we find 
\begin{equation}
 ds=\sqrt{\langle\Delta H_{U}^2\rangle}dt.
  \label{eq:Buress_metric2}
\end{equation}
where $\Delta H_{U}^2(t) = \langle H_{U}^2(t) \rangle - \langle H_{U}(t) \rangle^2$. Integrating this line element over the charging window $[t_i, t_f]$ yields the total dynamical path length $\mathcal{L} = \int_{t_i}^{t_f} \sqrt{\langle \Delta H_{U}^2(t) \rangle} \, dt$ traced out by the state $|\psi(t)\rangle$ \cite{Taddei2013}.
According to the Mandelstam-Tamm formulation of the quantum speed limit, the shortest possible path between the initial state $|\psi(t_i)\rangle$ and final state $|\psi(t_f)\rangle$ is the geodesic arc on the state manifold, whose length is given by the geodesic distance
\begin{equation}
    \mathcal{L}_{\mathrm{geo}} = \arccos\left( |\langle \psi(t_i) | \psi(t_f) \rangle| \right) .
    \label{eq:geodesic_length}
\end{equation}
Because the geodesic represents the absolute minimum distance connecting the two boundary states, the geometric path inequality $\mathcal{L} \ge \mathcal{L}_{\mathrm{geo}}$ naturally leads to the definition of the quantum geodesic efficiency
\begin{equation}
    \eta_{\mathrm{geo}} \equiv \frac{\mathcal{L}_{\mathrm{geo}}}{\mathcal{L}} = \frac{\arccos\left( |\langle \psi(t_i) | \psi(t_f) \rangle| \right)}{\int_{t_i}^{t_f} \sqrt{\langle \Delta H_{U}^2(t) \rangle} \, dt} .
    \label{eq:geodesic_eff_general}
\end{equation}
For our driven-dissipative cavity initialized in the vacuum state $|\psi(t_i)\rangle = |0\rangle$, the system remains in a pure coherent state $|\alpha(t)\rangle$ throughout the charging protocol. 
The overlap with the final state $|\alpha(t_f)\rangle$ is uniquely governed by the displacement amplitude, $|\langle 0 | \alpha(t_f) \rangle| = \exp\left(-\frac{1}{2}|\alpha(t_f)|^2\right)$ and $\sqrt{\langle \Delta H_{U}^2(t) \rangle} = |\dot{\alpha}(t)|$. Under these conditions, the quantum geodesic efficiency evaluates to the explicit expression
\begin{equation}
    \eta_{\mathrm{geo}} = \frac{\arccos\left(e^{-\frac{1}{2}|\alpha(t_f)|^2}\right)}{\int_0^{t_f} |\dot{\alpha}(t)|\,dt}\le 1 .
    \label{eq:geodesic_eff_coherent}
\end{equation}
The  inequality 
$\int|\dot\alpha|\,dt\ge\big|\int\dot\alpha\,dt\big|=|\Delta\alpha|$ where $\Delta\alpha=(\alpha_f-\alpha_i)$
gives
\begin{equation}
  \eta_{\rm geo}\le\eta_{\rm max}\equiv
  \frac{\arccos\!\big(e^{-|\Delta\alpha|^2/2}\big)}{|\Delta\alpha|}.
  \label{eq:S_bound}
\end{equation}
%For $\Omega_{\rm tot}=\Omega+\dot\Omega/\gamma_{\rm eff}$ with $\Omega(0)=0$,  Eq.~(\ref{eq:S_alpha}) is solved exactly by
%   \begin{equation}
%   \alpha_{\rm CD}(t)=\frac{i\lambda}{\gamma_{\rm eff}}\Omega(t).
 %  \label{eq:S_alphaCD}
%  \end{equation}
%   Since $\Omega(t)$ rises monotonically from $0$ to $\Omega_0$, $\dot\alpha$ has a constant phase, the path is a straight line of length $|\Delta\alpha|=a\equiv\lambda\Omega_0/|\gamma_{\rm eff}|$, and the bound (\ref{eq:S_bound}) is saturated:
%  \begin{equation}
%   \eta_{\rm geo}^{\rm CD}=\frac{\arccos(e^{-a^2/2})}{a}.
%  \end{equation}
%مقادیر تکمیل شود
%  For the parameters of Fig.~\ref{fig:cost}, $a=0.047$, so
%  $\eta_{\rm geo}^{\rm CD}=0.99982$ independent of $t_f$.

%  \subsection*{X.5 Path without CD}

%  For $\Omega_{\rm tot}=\Omega$, integrating Eq.~(\ref{eq:S_alpha}) by parts gives
%  \begin{equation}
%  \alpha(t)=\frac{i\lambda}{\gamma_{\rm eff}}\Omega(t)
%  -\frac{i\lambda}{\gamma_{\rm eff}}\int_0^t e^{-\gamma_{\rm eff}(t-s)}\dot\Omega(s)\,ds .
%  \end{equation}
%  The first term is the CD path. The second is a transient that oscillates at $\omega_{\rm eff}$ and decays at $\kappa/2$. It is the lag of the cavity behind the drive. It moves $\alpha(t)$ off the straight line, which increases $\int|\dot\alpha|\,dt$ and lowers $\eta_{\rm geo}$. The transient issmall when $\dot\Omega$ varies slowly compared with $\omega_{\rm eff}$,  and $\eta_{\rm geo}\to1$.

%%%%%%%%%%%%%%%%%%%%%%%%%%%%%%%%%%%%%%%%%
%000000000000000000000000000000000000000000000000000000000

%ادیت توسط چت جی پی تی
\subsection{Derivation of the Counterdiabatic Hamiltonian from the Liouvillian}
\label{app:liouvillian_cd}

We derive the auxiliary CD Hamiltonian directly from the time-dependent
Liouvillian of the driven-dissipative cavity.
% The construction uses a time-dependent displacement that maps the instantaneous steady state of thereference dynamics onto the vacuum. In the displaced frame, the referenceLiouvillian becomes time independent, while the time dependence of thetransformation generates an inertial Hamiltonian. An auxiliary Hamiltonianis then constructed to cancel this inertial contribution exactly. Thisapproach provides a state-tracking construction for the present open systemand is related to the Liouvillian formulation of adiabatic dynamics inRef.~[36]. In the closed-system limit, it recovers the standardDemirplak--Rice--Berry counterdiabatic Hamiltonian for displaced oscillatoreigenstates [17,18].
We first establish the transformation rules for closed and open systems \cite{Sarandy2005}.
Consider a closed system governed by
\begin{equation}
 i\frac{d}{dt}|\psi(t)\rangle=\hat H(t)|\psi(t)\rangle.
 \label{eq:app_cd_schrodinger}
\end{equation}
Introducing a differentiable unitary transformation $\hat U(t)$ and defining
$|\psi_d(t)\rangle=\hat U^\dagger(t)|\psi(t)\rangle$, differentiation gives
\begin{equation}
 i\frac{d}{dt}|\psi_d(t)\rangle=\hat H_d(t)|\psi_d(t)\rangle,
\end{equation}
where
\begin{equation}
 \hat H_d(t)=\hat U^\dagger(t)\hat H(t)\hat U(t)
 -i\hat U^\dagger(t)\dot{\hat U}(t).
 \label{eq:app_cd_closed_transform}
\end{equation}
%The second term is the inertial contribution arising from the time dependence of the frame.

For an open system described by the master equation
$\dot{\hat\rho}(t)=\mathcal L(t)\hat\rho(t)$, the transformed density operator
is $\hat\rho_d(t)=\hat U^\dagger(t)\hat\rho(t)\hat U(t)$. Differentiating and
using $\dot{\hat U}^\dagger\hat U=-\hat U^\dagger\dot{\hat U}$ yields
\begin{equation}
 \dot{\hat\rho}_d=\mathcal L_d(t)\hat\rho_d
 -[\hat U^\dagger\dot{\hat U},\hat\rho_d],
 \label{eq:app_cd_open_transform}
\end{equation}
where the transformed Liouvillian is defined by
\begin{equation}
 \mathcal L_d(t)\hat\rho_d=\hat U^\dagger(t)
 \mathcal L(t)[\hat U(t)\hat\rho_d\hat U^\dagger(t)]\hat U(t).
 \label{eq:app_cd_transformed_liouvillian}
\end{equation}
The additional commutator in Eq.~\eqref{eq:app_cd_open_transform} can be
written as
\begin{equation}
 -[\hat U^\dagger\dot{\hat U},\hat\rho_d]
 =-i[\hat H_I,\hat\rho_d],\qquad
 \hat H_I=-i\hat U^\dagger\dot{\hat U}.
 \label{eq:app_cd_inertial_general}
\end{equation}
Thus, the same inertial operator arises in both closed and open systems,
appearing in the transformed Hamiltonian in the former case and in the
commutator contribution to the transformed Liouvillian in the latter.

We now apply this transformation to the reduced cavity dynamics. To allow for
both in-phase and quadrature control, we write the reference Hamiltonian as
\begin{equation}
 \hat H_0(t)=\omega_{\mathrm{eff}}\hat a^\dagger\hat a
 -\lambda[\Omega(t)\hat a^\dagger+\Omega^*(t)\hat a],
 \label{eq:app_cd_ref_H}
\end{equation}
where $\lambda$ is real, as in \eqref{eq:lambda}. For a real reference
envelope this reduces to Eq.~\eqref{eq:Heff_final}. The reference master
equation is
\begin{align}
 \dot{\hat\rho}&=\mathcal L_0(t)\hat\rho,\nonumber\\
 \mathcal L_0(t)\hat\rho&=-i[\hat H_0(t),\hat\rho]
 +\kappa\mathcal D[\hat a]\hat\rho,
 \label{eq:app_cd_ref_L}
\end{align}
with $\mathcal D[\hat a]\hat\rho=\hat a\hat\rho\hat a^\dagger
-\tfrac12\{\hat a^\dagger\hat a,\hat\rho\}$. For $\kappa>0$, the instantaneous
steady state of the reference Liouvillian is the coherent state with amplitude
$
 \alpha_{\mathrm{ad}}(t)=\frac{i\lambda\Omega(t)}{\gamma_{\mathrm{eff}}}
$.
The parameters $\lambda$ and $\gamma_{\mathrm{eff}}$ are taken to be
constant during the charging interval.

Introduce the displacement operator and its corresponding superoperator,
\begin{align}
 \hat D(\alpha)&=\exp(\alpha\hat a^\dagger-\alpha^*\hat a),\nonumber\\
 \mathcal U_D(\alpha)\hat X&=\hat D(\alpha)\hat X\hat D^\dagger(\alpha),
\end{align}
and define $\hat\rho_d=\hat D^\dagger(\alpha)\hat\rho\hat D(\alpha)$.
The displacement identities are
\begin{equation}
 \hat D^\dagger\hat a\hat D=\hat a+\alpha,\qquad
 \hat D^\dagger\hat a^\dagger\hat D=\hat a^\dagger+\alpha^*.
 \label{eq:app_cd_shift}
\end{equation}
Under the instantaneous transformation, the Hamiltonian part becomes
\begin{align}
 \hat D^\dagger\hat H_0\hat D
 &=\omega_{\mathrm{eff}}\hat a^\dagger\hat a
 +[(\gamma_{\mathrm{eff}}\alpha-\lambda\Omega)\hat a^\dagger
 +\mathrm{H.c.}].
 \label{eq:app_cd_shift_H}
\end{align}
The dissipator transforms independently as
\begin{align}
 \mathcal D[\hat a+\alpha]\hat\rho_d
 &=\mathcal D[\hat a]\hat\rho_d
 +\frac{\alpha^*}{2}[\hat a,\hat\rho_d]
 -\frac{\alpha}{2}[\hat a^\dagger,\hat\rho_d]\nonumber\\
 &=\mathcal D[\hat a]\hat\rho_d
 -i\left[\frac{i}{2}(\alpha^*\hat a-\alpha\hat a^\dagger),\hat\rho_d\right].
 \label{eq:app_cd_shift_diss}
\end{align}
Combining Eqs.~\eqref{eq:app_cd_shift_H} and~\eqref{eq:app_cd_shift_diss},
the instantaneous transformed Liouvillian
$\mathcal L_J=\mathcal U_D^{-1}\mathcal L_0\mathcal U_D$ is
\begin{align}
 \mathcal L_J\hat\rho_d&=-i[\hat H_J,\hat\rho_d]
 +\kappa\mathcal D[\hat a]\hat\rho_d,\nonumber\\
 \hat H_J&=\omega_{\mathrm{eff}}\hat a^\dagger\hat a
 +\left[\left((\omega_{\mathrm{eff}}-i\kappa/2)\alpha
 -\lambda\Omega\right)\hat a^\dagger+\mathrm{H.c.}\right].
 \label{eq:app_cd_combined_J}
\end{align}
Choosing $\alpha=\alpha_{\mathrm{ad}}$
from Eq.~\eqref{eq:alpha_ad_instantaneous} cancels all linear terms in  \eqref {eq:app_cd_combined_J}.  Hence the reference Liouvillian reduces to the
time-independent damped-oscillator generator
\begin{equation}
 \mathcal L_J\hat\rho_d=-i[\omega_{\mathrm{eff}}\hat a^\dagger\hat a,
 \hat\rho_d]+\kappa\mathcal D[\hat a]\hat\rho_d.
 \label{eq:app_cd_J}
\end{equation}
Its stationary state is the vacuum, $\mathcal L_J(|0\rangle\langle0|)=0$.
The time dependence of the displacement has not yet been included in
Eq.~\eqref{eq:app_cd_J}, it is accounted for by the inertial term in
Eq.~\eqref{eq:app_cd_open_transform}.

For the time-dependent displacement $\hat D(\alpha_{\mathrm{ad}}(t))$,
we use the identity
\begin{align}
 \hat D^\dagger\dot{\hat D}
 &=\dot\alpha_{\mathrm{ad}}\hat a^\dagger
 -\dot\alpha_{\mathrm{ad}}^*\hat a+q(t),\nonumber\\
 q(t)&=\tfrac12(\alpha_{\mathrm{ad}}^*\dot\alpha_{\mathrm{ad}}
 -\alpha_{\mathrm{ad}}\dot\alpha_{\mathrm{ad}}^*).
 \label{eq:app_cd_D_derivative}
\end{align}
Here $q(t)$ is purely imaginary, so that $-iq(t)$ is real. This identity
follows from $\hat D=e^{\hat A}$ with
$\hat A=\alpha_{\mathrm{ad}}\hat a^\dagger-\alpha_{\mathrm{ad}}^*\hat a$:
$\hat D^\dagger\dot{\hat D}=\dot{\hat A}-[\hat A,\dot{\hat A}]/2$.
The inertial Hamiltonian is therefore
\begin{equation}
 \hat H_I=-i\left(\dot\alpha_{\mathrm{ad}}\hat a^\dagger
 -\dot\alpha_{\mathrm{ad}}^*\hat a\right)-iq(t).
 \label{eq:app_cd_HI}
\end{equation}
In the displaced frame, the full reference equation of motion is
\begin{equation}
 \dot{\hat\rho}_d=\mathcal L_J\hat\rho_d-i[\hat H_I,\hat\rho_d].
 \label{eq:app_cd_inertial_eom}
\end{equation}
The operator part of $\hat H_I$ drives the system away from the vacuum and
is responsible for the lag relative to the moving reference steady state.
No assumption about the Jordan canonical form of $\mathcal L_J$ is required
for this state-tracking construction.

To remove this inertial contribution, we add an auxiliary Hamiltonian
$\hat H_{\mathrm{CD}}$ to the original reference Hamiltonian, requiring
\begin{equation}
 \hat D^\dagger\hat H_{\mathrm{CD}}\hat D=-\hat H_I,
 \label{eq:app_cd_cancel}
\end{equation}
where equality up to a scalar multiple of the identity is sufficient.
An exact choice including its scalar gauge term is
\begin{equation}
 \hat H_{\mathrm{CD}}=i\dot{\hat D}\hat D^\dagger.
 \label{eq:app_cd_exact_H}
\end{equation}
Accordingly, after omitting scalar terms that do not influence the master
equation, the required CD Hamiltonian can be written as
\begin{align}
 \hat H_{\mathrm{CD}}(t)&=i(\dot\alpha_{\mathrm{ad}}\hat a^\dagger
 -\dot\alpha_{\mathrm{ad}}^*\hat a)\nonumber\\
 &=-\lambda[\Omega_{\mathrm{CD}}(t)\hat a^\dagger
 +\Omega_{\mathrm{CD}}^*(t)\hat a],
 \label{eq:app_cd_final_H}
\end{align}
where substitution of Eq.~\eqref{eq:alpha_ad_instantaneous} yields
\begin{equation}
\Omega_{\mathrm{CD}}(t)=\frac{\dot\Omega(t)}{\gamma_{\mathrm{eff}}}.
 \label{eq:app_cd_field}
\end{equation}
This convention treats $\Omega_{\mathrm{CD}}$ as the auxiliary field applied
to the charger, so that the effective cavity drive is
$\lambda\Omega_{\mathrm{CD}}$. The total Hamiltonian retains the form of
Eq.~\eqref{eq:app_cd_ref_H}, with $\Omega$ replaced by
$\Omega_{\mathrm{tot}}=\Omega+\Omega_{\mathrm{CD}}$.

With this auxiliary Hamiltonian, the inertial commutator is canceled, and the
transformed master equation becomes
\begin{equation}
 \dot{\hat\rho}_d=\mathcal L_J\hat\rho_d.
\end{equation}
For $\Omega(0)=0$ and the initial vacuum state, the transformed density
operator remains $\hat\rho_d(t)=|0\rangle\langle0|$. Transforming back gives
\begin{equation}
 \hat\rho(t)=\hat D(\alpha_{\mathrm{ad}}(t))|0\rangle\langle0|
 \hat D^\dagger(\alpha_{\mathrm{ad}}(t))
 =|\alpha_{\mathrm{ad}}(t)\rangle\langle\alpha_{\mathrm{ad}}(t)|.
 \label{eq:app_cd_tracking}
\end{equation}
Thus, the CD field enforces exact tracking of the reference coherent
steady-state trajectory within the effective model. The cavity dissipator
with rate $\kappa$ remains present.  It is the inertial contribution, rather
than the environmental dissipation itself, that is canceled.

Finally, we establish the connection with conventional transitionless
quantum driving. In the closed-system limit $\kappa\to0$, for fixed nonzero
$\omega_{\mathrm{eff}}$, we have
\begin{equation}
 \gamma_{\mathrm{eff}}\to i\omega_{\mathrm{eff}},\qquad
 \alpha_{\mathrm{ad}}\to\frac{\lambda\Omega}{\omega_{\mathrm{eff}}},\qquad
 \Omega_{\mathrm{CD}}\to-\frac{i\dot\Omega}{\omega_{\mathrm{eff}}}.
\end{equation}
Consequently, the CD Hamiltonian becomes, up to a scalar term,
\begin{equation}
 \hat H_{\mathrm{CD}}\to\frac{i\lambda}{\omega_{\mathrm{eff}}}
 [\dot\Omega\hat a^\dagger-\dot\Omega^*\hat a].
 \label{eq:app_cd_closed_limit}
\end{equation}
For $\kappa=0$, the reference Hamiltonian can be expressed as
\begin{equation}
 \hat H_0(t)=\omega_{\mathrm{eff}}(\hat a^\dagger-\alpha_{\mathrm{ad}}^*)
 (\hat a-\alpha_{\mathrm{ad}})-\omega_{\mathrm{eff}}|\alpha_{\mathrm{ad}}|^2,
\end{equation}
where $\alpha_{\mathrm{ad}}=\lambda\Omega/\omega_{\mathrm{eff}}$.
Its instantaneous eigenstates are the displaced Fock states
\begin{equation}
 |n;t\rangle=\hat D(\alpha_{\mathrm{ad}}(t))|n\rangle.
\end{equation}
The standard Berry counterdiabatic Hamiltonian [18] has the form
\begin{equation}
 \hat H_{\mathrm{CD}}^{\mathrm{Berry}}
 =i\sum_n\left(|\partial_t n;t\rangle\langle n;t|
 -\langle n;t|\partial_t n;t\rangle|n;t\rangle\langle n;t|\right).
 \label{eq:app_cd_berry}
\end{equation}
Completeness of the displaced Fock basis gives
\begin{equation}
 \sum_n|\partial_t n;t\rangle\langle n;t|=\dot{\hat D}\hat D^\dagger.
\end{equation}
Furthermore, Eq.~\eqref{eq:app_cd_D_derivative} implies
$\langle n;t|\partial_t n;t\rangle=q(t)$ because the expectation values
of $\hat a$ and $\hat a^\dagger$ vanish in number states. Thus the diagonal
subtraction is a scalar, independent of $n$, and
\begin{align}
 \hat H_{\mathrm{CD}}^{\mathrm{Berry}}
 &=i\dot{\hat D}\hat D^\dagger-iq(t)\hat{\rm I}\nonumber\\
 &=i(\dot\alpha_{\mathrm{ad}}\hat a^\dagger
 -\dot\alpha_{\mathrm{ad}}^*\hat a)-2iq(t)\hat{\rm I}.
 \label{eq:app_cd_berry_final}
\end{align}
The last equality follows from
$\dot{\hat D}\hat D^\dagger=\dot\alpha_{\mathrm{ad}}\hat a^\dagger
-\dot\alpha_{\mathrm{ad}}^*\hat a-q(t)\hat{\rm I}$.
For the real reference envelope used in the main text, the displacement has
a fixed phase and $q(t)=0$. More generally, the scalar term has no effect on
the density-operator dynamics. Therefore, up to a physically irrelevant scalar gauge term,
\begin{equation}
 \lim_{\kappa\to0}\hat H_{\mathrm{CD}}^{\mathrm{open}}
 =\hat H_{\mathrm{CD}}^{\mathrm{Berry}},\qquad \omega_{\mathrm{eff}}\ne0.
\end{equation}
This establishes the continuous connection between the open-system
steady-state tracking protocol and conventional transitionless driving of
the displaced oscillator. The simultaneous resonant and zero-damping limit
is excluded because the reference stationary displacement then becomes
singular for nonzero drive.

\subsection { Lower bound on the counterdiabatic control cost–dissipation product}
\label{Lower bound}
To establish a lower bound on the trade-off between CD control cost and dissipative losses, we consider the control cost  defined in Eq.~\eqref{eq:total_fluence_cost} for $\Omega_{{\rm CD}}$ defined in Eq.~\eqref {eq:cd_field_exact} as
\[
C_{\mathrm{CD}}=\frac{1}{\lambda^2}
\int_0^{t_f}|\dot{\alpha}(t)|^2\,dt,
\]
 The energy dissipated through photon leakage during the charging process is given by
\[
E_{\mathrm{diss}}=\kappa\omega_a
\int_0^{t_f}|\alpha(t)|^2\,dt,
\]
Applying the Cauchy–Schwarz inequality to the two integrals yields

\[
\begin{aligned}
C_{\mathrm{CD}}E_{\mathrm{diss}}
&=\frac{\kappa\omega_a}{\lambda^2}
\left(\int_0^{t_f}|\dot{\alpha}(t)|^2dt\right)
\left(\int_0^{t_f}|\alpha(t)|^2dt\right)\\
&\geq \frac{\kappa\omega_a}{\lambda^2}
\left(\int_0^{t_f}|\alpha(t)|\,|\dot{\alpha}(t)|dt\right)^2.
\end{aligned}
\]

Next, using

\[
\frac{d}{dt}|\alpha(t)|^2
=2\operatorname{Re}\left[\alpha^*(t)\dot{\alpha}(t)\right],
\]
and the inequality
\[
|\alpha(t)|\,|\dot{\alpha}(t)|
\geq \operatorname{Re}\left[\alpha^*(t)\dot{\alpha}(t)\right],
\]
we obtain
\[
\begin{aligned}
\int_0^{t_f}|\alpha(t)|\,|\dot{\alpha}(t)|dt
&\geq \frac{1}{2}\int_0^{t_f}
\frac{d}{dt}|\alpha(t)|^2dt\\
&=\frac{1}{2}\left[|\alpha(t_f)|^2-
|\alpha(0)|^2\right].
\end{aligned}
\]

For a battery initially in the vacuum state, \(\alpha(0)=0\), and therefore

\[
\int_0^{t_f}|\alpha(t)|\,|\dot{\alpha}(t)|dt
\geq \frac{|\alpha_f|^2}{2},
\]

where \(\alpha_f\equiv\alpha(t_f)\). Substituting this result into the preceding expression gives

\[
C_{\mathrm{CD}}E_{\mathrm{diss}}
\geq \frac{\kappa\omega_a}{4\lambda^2}
|\alpha_f|^4.
\]

Finally, since the final stored energy is
\(E_f=\omega_a|\alpha_f|^2\), the resulting bound can be expressed as

\begin{equation}
C_{\mathrm{CD}}E_{\mathrm{diss}}
\geq
\frac{\kappa E_f^2}
{4\omega_a\lambda^2}.
\label{A1}
\end{equation}

Eq.~\eqref{A1} establishes a lower bound on the product of the CD control cost and dissipative losses for a prescribed final battery energy. Consequently, these two quantities cannot be simultaneously reduced arbitrarily while maintaining the same charging target. The bound is independent of the specific temporal profile of the charging protocol and therefore provides a general constraint on the charging dynamics within the considered model.

Equality in the Cauchy–Schwarz inequality requires \(|\dot{\alpha}(t)|\) to be proportional to \(|\alpha(t)|\) throughout the charging interval. In addition, equality in the second inequality requires the evolution of \(\alpha(t)\) to follow a purely radial, nondecreasing trajectory in the complex-amplitude plane. These conditions are generally not simultaneously attainable for a nonzero final amplitude when the initial condition is \(\alpha(0)=0\) and the charging time is finite. Thus, Eq. (A1) should be interpreted as a general lower bound rather than an exact minimum for a particular charging protocol.
\section{Acknowledgment }

\end{document}